\documentclass[12pt]{iopart}

\usepackage{float}
\usepackage{graphicx}% Include figure files
\usepackage{dcolumn}% Align table columns on decimal point
\usepackage{bm}% bold math
\usepackage{hyperref}
\hypersetup{
	colorlinks=true,
	citecolor=blue,
	linkcolor=blue,
	urlcolor=blue
}
\usepackage{setspace}
\usepackage{caption}
\usepackage{iopams}
\usepackage{cite}
\begin{document}

\title[]{Opinions with few disciples can win in the dynamical directed networks: an evolutionary game perspective}
\author{Yakun Wang and Bin Wu\thanks{*Corresponding author}}
%\author{Bin Wu\thanks{*Corresponding author}}
\ead{bin.wu@bupt.edu.en}
\address{School of Science, Beijing University of Posts and Telecommunications, Beijing 100876, China.}

%\author{Content \& Services Team}
%
%\address{IOP Publishing, Temple Circus, Temple Way, Bristol BS1 6HG, UK}
%\ead{submissions@iop.org}
%\vspace{10pt}
%\begin{indented}
%\item[]August 2017
%\end{indented}

\begin{abstract}
The voter model on networks is crucial to understand opinion formation. Uni-directional social interactions are ubiquitous in real social networks whereas undirected interactions are intensively studied. We establish a voter model on a dynamical directed network. We show that the opinion invasion is captured by a replicator equation of an emergent four-player two-strategy game, and the average in(out)-degree for the two opinions is fully captured by an emergent three-player two-strategy game. Interestingly, it is shown that the difference between the two emergent games arises from the uni-directionality of the network. The difference implies that the opinion with a small number of disciples can take over the population for in-group bias, provided that the network is directed. Our work makes an explicit connection between opinion dynamics and evolutionary games.
\end{abstract}

%
% Uncomment for keywords
%\vspace{2pc}
%\noindent{\it Keywords}: XXXXXX, YYYYYYYY, ZZZZZZZZZ
%
% Uncomment for Submitted to journal title message
%\submitto{\JPA}
%
% Uncomment if a separate title page is required
%\maketitle
%
% For two-column output uncomment the next line and choose [10pt] rather than [12pt] in the \documentclass declaration
%\ioptwocol
%

\section{\label{sec:level1}Introduction}
Opinion dynamics have become attractive in diverse disciplines, such as statistical physics, control theory and system science \cite{Castellano_2009, Opinion_Dynamics,doi:10.1137/11082751X,7219413,8521596,PhysRevX.3.021004}. Two main topics of opinion dynamics are how opinions reach a consensus and how opinions coexist for a long time. The voter model is one of the classical models \cite{PhysRevLett.94.178701, PhysRevLett.112.158701, PhysRevE.80.041129}. It is a discrete opinion dynamics model in which an individual adopts an opinion with a probability proportional to the fraction of that opinion in its neighborhood. Besides opinion dynamics, the voter model has various applications in many fields, such as epidemic spreading \cite{Pinto_2011}, catalytic reactions in chemistry \cite{PhysRevE.53.R3009} and prey-predator interaction in biology \cite{PhysRevE.70.012901}.

Individual interactions in opinion dynamics are typically captured by networks. The real-world networks are dynamical, rather than static \cite{PhysRevLett.97.258103, PhysRevLett.95.098104, PERC2010109, Nature1, WEI2019109978, WuEvolving, linking, Social, Bridging}. The researches on the co-evolutionary dynamics of opinions and networks have been well thorough \cite{PhysRevE.100.042303, PhysRevE.84.046111, PhysRevE.83.035101, KUROKAWA201913}. A simple model with a single parameter controlling the balance of the two dynamics is built to investigate the opinion formation \cite{PhysRevE.74.056108}. The modified model exhibits complicated topological behaviors via introducing heterophily \cite{PhysRevE.78.016103}. One individual can rewire to an individual chosen at random from those with the same opinion or from the whole network. The rewire-to-same and rewire-to-random models have different phase transitions \cite{doi:10.1073/pnas.1200709109}. Master equation approximation, pair approximation and heterogeneous mean-field are well-known approaches to capture the opinion dynamics on the networks \cite{PhysRevLett.94.178701, Vazquez_2008, PhysRevE.79.046104, Peralta_2020}. But all of these works explicitly assume that the networks are bi-directional.

Unidirectional social interactions are ubiquitous in the real world. For example, a user follows another user on Twitter based on a common interest, and this following relationship is asymmetric \cite{STEPHENS201587}: Sally enjoys Pilates, so she follows the blogger Jessica, who teaches Pilates online. But Jessica does not follow Sally. In the US National Longitudinal Study of Adolescent Health (the “AddHealth” study), high school students were asked to identify their friends within the school. More than half of the friendships are found to be unidirectional. Lisa considering Cindy to be her friend does not imply that Cindy considers Lisa to be her friend \cite{doi:10.1073/pnas.0610537104}. A large number of biological systems also have unidirectional interactions. For example in a wolf pack, wolves in general are subservient to the alpha wolf and their socialization is strictly one-way \cite{1996A}. Directed dynamic networks are also widely present in the field of engineering \cite{HAMDIPOOR20213127, HAN201599}. We concentrate on the unidirectional nature of the network \cite{directed, doi:10.1073/pnas.2113468118} besides the dynamic nature of the social network.

In this paper, we establish a voter model on a dynamical directed network \cite{doi:10.1073/pnas.97.16.9340, Gross_2007}. Each node in the network represents an individual, and each directed link represents a directed social relationship. We are to address two questions, i.e., fate of opinions and transient topology. It is found that the fate of opinions is captured by an \emph{emergent} four-player two-strategy game. The expectation of in(out)-degree for the two opinions is captured by an \emph{emergent} three-player two-strategy game. The two emergent games are typically different for directed networks, which facilitates us to explain some counterintuitive phenomena.

%%%%%%%%%%%%%%%%%%%%%%%%%%%%%%%%%%%

\section{Model}
\begin{figure}
	\centering
	\includegraphics[scale=0.32]{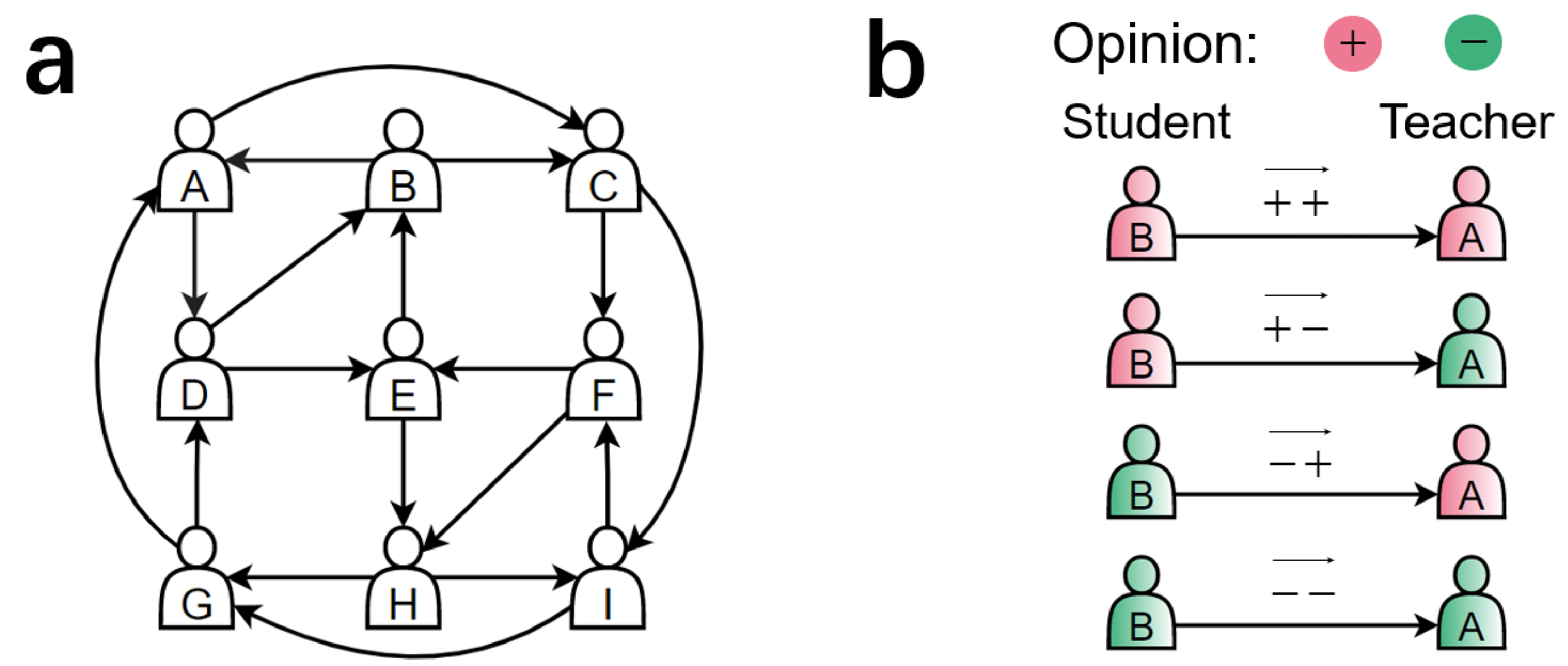}% Here is how to import EPS art
	\caption{\label{tu1} \textbf{Uni-directional social interactions.} (a) This is a regular directed network. There are nine individuals and each individual has two incoming and two outgoing edges, i.e., $N = 9$ and $L = 2$. (b) There are two types of opinions and four types of edges, namely $\protect\overrightarrow { +  + }, \protect\overrightarrow { +  - }, \protect\overrightarrow { -  + }, \protect\overrightarrow { -  - } $. A directed edge connects $A$ and $B$, which implies that $B$ as a student can learn from $A$ as a teacher.}
\end{figure}

Initially, the whole population of size $N$ are situated on nodes of a regular directed graph. Each node has $L$ incoming edges and $L$ outgoing edges, as shown in Fig. \hyperref[tu1]{1(a)}. The total number of directed links is thus $NL$. We assume that $N \gg L$. It implies that each individual has a limited number of neighbors compared with the population size which is ubiquitous in social networks. There are two opinions, denoted as $+$ and $-$, respectively. Each individual holds one type of opinion and we denote $\overrightarrow {XY} $ as the type of the directed link, where $\overrightarrow {XY}  \in \left\{ {\overrightarrow { +  + } ,\overrightarrow { +  - } ,\overrightarrow { -  + } ,\overrightarrow { -  - } } \right\} \buildrel \Delta \over = S$.

Here we propose a voter model on the evolving directed network. In the network, we define the direction of “learning”: for example, if node $B$ points to node $A$, it implies that $B$ unilaterally learns from $A$ and $A$ does not learn from $B$. In other words, the source node plays the role of a student to learn the target node who plays the role of a teacher, as shown in Fig. \hyperref[tu1]{1(b)}. For a node, it has a student set and a teacher set. The student set is composed of the source nodes on the edges that flow into the node, and the teacher set is composed of the target nodes on the edges that flow out from the node.

\begin{figure}
	\centering
	\includegraphics[scale=0.27]{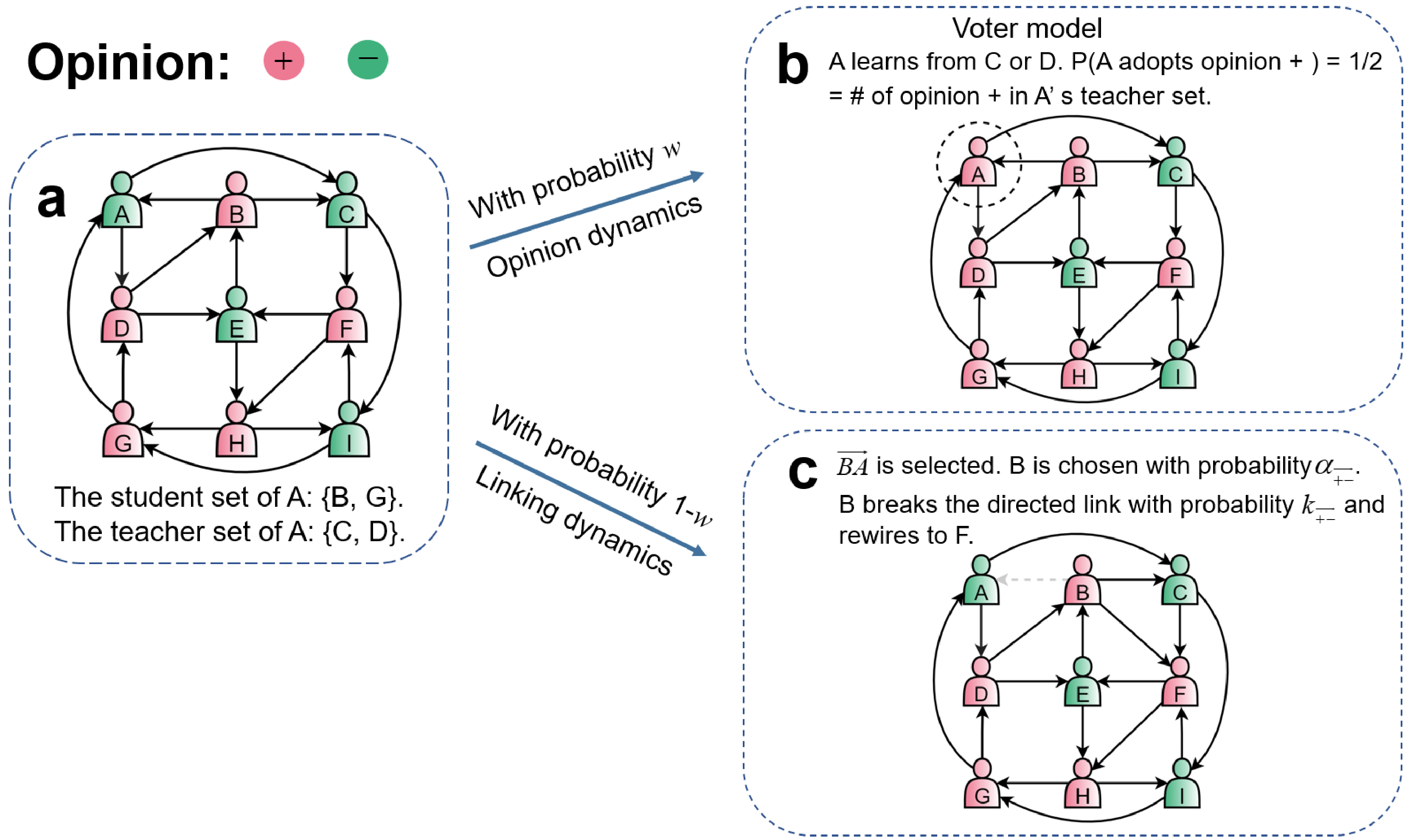}% Here is how to import EPS art
	\caption{\label{tu2} \textbf{Coevolutionary dynamics of opinions and directed social relationships.} (a) A population is described by a regular directed network. For example, the student set of $A$ is $\left\{ {B,G} \right\}$ and the teacher set of $A$ is $\left\{ {C,D} \right\}$. (b) With probability $w$, opinion update happens. In this case, an individual is randomly chosen to update its opinion. It learns from its teachers who are the target nodes. The probability of the focal individual adopting the opinion $ \pm $ is proportional to the number of its teachers with opinion $ \pm $. For example, suppose $A$ is selected and its teachers are $C$ and $D$. Then $A$ adopts opinion $+$ with probability 1/2. If the selected node has no outgoing edges, i.e., it has no teacher to learn from, then it will remain the original opinion. (c) With probability $1-w$, the linking dynamics happens. Firstly, a directed link is selected randomly. Secondly, the source or target node of the directed link is chosen based on the respective pre-defined probability. Thirdly, the directed link breaks with a pre-defined probability depending on its type. If the student(teacher) is selected and the directed link is broken, then the node reconnects with the outgoing(incoming) edge to a random node that is neither in its current teacher set nor in its current student set. For example, the directed link $\protect\overrightarrow {BA} $ is selected. With probability ${\alpha _{\scriptsize\protect\overrightarrow { +  - } }}$, the source node $B$ is selected and the directed link is broken with probability ${k_{\scriptsize\protect\overrightarrow { +  - } }}$. Then the node randomly chooses an individual which is neither in $B$'s current teacher set nor in $B$'s current student set ($F$,$G$,$H$ or $I$). Otherwise, with probability $\beta_{\scriptsize\protect\overrightarrow { +  - } }$, the target node $A$ is selected, and then it also randomly chooses an individual which is neither in $A$'s current teacher set nor in $A$'s current student set ($E$,$F$,$H$ or $I$). Suppose $B$ is selected and breaks the directed link $\protect\overrightarrow {BA} $. Finally, $B$ finds a new teacher $F$ and reconnects to $F$ with a directed link. }
\end{figure}

Each individual has an opportunity to either update its opinion with probability $w$ or update its link with probability $1 - w$ at each time step, which is shown in Fig. \hyperref[tu2]{2}. When $w = 1$, the social links between individuals are invariant, i.e., individuals only update their opinions. It refers to the opinion dynamics on a static directed network \cite{PhysRevE.74.026114,PhysRevE.81.057103}. When $w = 0$, the social network evolves all the time whereas the fractions of opinions are constant.

For \textbf{\itshape{opinion dynamics}}, we focus on the voter model \cite{PhysRevX.3.021004}. An individual is randomly selected from the population. The probability that the selected individual adopts opinion $+$ is proportional to the number of teachers with opinion $+$ in its teacher set. In other words, the selected individual adopts opinion $+$ with probability ${Q_{\scriptsize +}}/\left( {{Q_{\scriptsize +}} + {Q_{\scriptsize -}}} \right)$, where ${Q_ {\scriptsize \pm} }$ refers to the number of its teachers whose opinion is $ \pm $. It is notable that if the teacher set of the selected node is empty, then the individual has no teachers to learn from and keeps the opinion.

For \textbf{\itshape{linking dynamics}}, our model focuses on the updating of directed links. The whole network is adjusted by at most one directed link at each time step. There are three steps as follows.

(i) \emph{Selecting a directed link.} A directed link $\overrightarrow {XY} $ is randomly selected from all the directed links. The directed link $\overrightarrow {XY} $ corresponds to the student $X$ and the teacher $Y$, where $\overrightarrow {XY}  \in S$.

(ii) \emph{Selecting $X$ or $Y$.} $X$ is selected with probability ${\alpha _{\scriptsize\overrightarrow {XY} }}$, where $0<{\alpha_{\scriptsize\overrightarrow {XY} }}<1$. Otherwise $Y$ is selected with probability ${\beta_{\scriptsize\overrightarrow {XY} }}$. We have ${\alpha _{\scriptsize\overrightarrow {XY} }} + {\beta_{\scriptsize\overrightarrow {XY} }} = 1$.

(iii) \emph{Breaking the directed link.} The $\overrightarrow {XY} $ breaks off with a pre-defined probability ${k_{\scriptsize\overrightarrow{XY}}}$, where $0<{k_{\scriptsize\overrightarrow {XY} }}<1$. It implies that if the student $X$(teacher $Y$) is selected, then $X$($Y$) would like to break the directed link with probability ${k_{\scriptsize\overrightarrow {XY} }}$ to change the current teacher(student).

(iv) \emph{Rewiring the node.} If student $X$ is selected and the $\overrightarrow {XY} $ is broken, then $X$ will find a new teacher who is neither in $X$'s current teacher set nor in $X$'s current student set. If the teacher $Y$ is selected and the $\overrightarrow {XY} $ is broken, then $Y$ will teach a new student who is neither in $Y$'s current teacher set nor in $Y$'s current student set.

Notably, the number of teachers in the entire population is constant, since the sum of out-degrees of all the nodes in the network keeps unchanged over time.

\section{Emergent games for the fate of opinions}
The voter model on the evolving network is a Markov chain with state $x {\scriptsize +}$, i.e., the fraction of opinion $+$ in the population. Thus, the state space is $\left\{ {0,1/N,2/N, \cdots ,1} \right\}$. State 0 and state 1 are absorbing states, which implies that all the individuals reach a consensus. We focus on $w \ll 1$. In this case, individuals prefer to adjust their social relationships rather than change their opinions. This is widespread in real social systems. For example, users on Twitter change their opinions much less frequently than adjust their followers \cite{KROSNICK1988240}. It leads to a time scale separation, that is, all the directed links are almost in the stationary regime when the opinion update occurs (see \hyperref[Appendix A]{Appendix A} for details).

For the evolutionary dynamics of opinions, $x_{\scriptsize +}$ either increases or decreases by $1 / N$ within a time step. For example, $x_{\scriptsize +}$ increases by $1 / N$ if an individual who adopts opinion $-$ is selected with probability ${x_{\scriptsize -}} = 1 - {x_{\scriptsize +}}$, i.e., the fraction of opinion $-$ in the population. Then the focal individual with opinion $-$ learns from its teachers with opinion $+$. And it adopts opinion $+$ with a probability proportional to the number of its teachers with opinion $+$, i.e., ${{q{\pi _{\scriptsize\overrightarrow { -  + } }}} \mathord{\left/
		{\vphantom {{q{\pi _{\scriptsize\overrightarrow { -  + } }}} {\left( {q{\pi _{\scriptsize\overrightarrow { -  + } }} + q{\pi _{\scriptsize\overrightarrow { -  - } }}} \right)}}} \right.
		\kern-\nulldelimiterspace} {\left( {q{\pi _{\scriptsize\overrightarrow { -  + } }} + q{\pi _{\scriptsize\overrightarrow { -  - } }}} \right)}} = {{{\pi _{\scriptsize\overrightarrow { -  + } }}} \mathord{\left/
		{\vphantom {{{\pi _{\scriptsize\overrightarrow { -  + } }}} {\left( {q{\pi _{\scriptsize\overrightarrow { -  + } }} + {\pi _{\scriptsize\overrightarrow { -  - } }}} \right)}}} \right.
		\kern-\nulldelimiterspace} {\left( {{\pi _{\scriptsize\overrightarrow { -  + } }} + {\pi _{\scriptsize\overrightarrow { -  - } }}} \right)}}$.
Here $q$ is the average size of the teacher set captured by the average out-degree of the focal individual. Thus the transition probability that $x_{\scriptsize +}$ increases by $1 / N$ is
\begin{equation}
\label{eq.1}
T_{{x_{\scriptsize +}}}^ + = {x_{\scriptsize -}}\frac{{{\pi _{\scriptsize\overrightarrow { -  + } }}}}{{{\pi _{\scriptsize\overrightarrow { -  + } }} + {\pi _{\scriptsize\overrightarrow { -  - } }}}}.
\end{equation}

\noindent Similarly, the transition probability that $x {\scriptsize +}$ decreases by $1 / N$ is
\begin{equation}
\label{eq.2}
T_{{x_{\scriptsize +}}}^ - = {x_{\scriptsize +}}\frac{{{\pi _{\scriptsize\overrightarrow { +  - } }}}}{{{\pi _{\scriptsize\overrightarrow { +  + } }} + {\pi _{\scriptsize\overrightarrow { +  - } }}}}.
\end{equation}
\noindent The probability that $x {\scriptsize +}$ remains constant is $T_{{x_{\scriptsize +}}}^0 = 1 - T_{{x_{\scriptsize +}}}^ +  - T_{{x_{\scriptsize +}}}^ - $, since the each row sum of the transition probability matrix is unit one.

For large population size, i.e., $N \to +\infty $, the mean-field equation is given by ${\dot x_{\scriptsize +} } = T_{{x_{\scriptsize +}}}^ +  - T_{{x_{\scriptsize +}}}^ -$, capturing the evolution of the opinions. Taking Eqs. \hyperref[eq.1]{(1)}, \hyperref[eq.2]{(2)} yields that ${{\dot x}_ {\scriptsize +} } = {x_{\scriptsize +}}{x_{\scriptsize -}}$
$\left[ {} \right. {{{k_{\scriptsize\overrightarrow { -  - } }}\left( {{\alpha _{\scriptsize\overrightarrow { +  - } }}{\beta _{\scriptsize\overrightarrow { +  + } }}{x_{\scriptsize +}} + {\alpha _{\scriptsize\overrightarrow { -  - } }}{\beta _{\scriptsize\overrightarrow { +  - } }}{x_{\scriptsize -}}} \right)}	{A^{ - 1}}\left( {{x_{\scriptsize +}}} \right)} \!-\! {k_{\scriptsize\overrightarrow { +  + } }}\left( {{\alpha _{\scriptsize\overrightarrow { +  + } }}{\beta _{\scriptsize\overrightarrow { -  + } }}{x_{\scriptsize +}}} + {{\alpha _{\scriptsize\overrightarrow { -  + } }}{\beta _{\scriptsize\overrightarrow { -  - } }}{x_{\scriptsize -}}} \!\right){B^{ - 1}}\left( {{x_{\scriptsize +}}} \right) \left. {} \right]$, where
both $A\left( {{x_{\scriptsize +}}} \right) \ =\  {k_{\scriptsize\overrightarrow { -  - } }}{\alpha _{\scriptsize\overrightarrow { +  - } }}{\beta _{\scriptsize\overrightarrow { +  + } }}\, x_{\scriptsize +}^2 + \left[ {{\, \beta _{\scriptsize\overrightarrow { +  - } }}\left( {{k_{\scriptsize\overrightarrow { -  + } }}{\alpha _{\scriptsize\overrightarrow { +  + } }}} \right.} \right. + \left. {{k_{\scriptsize\overrightarrow { -  - } }}{\alpha _{\scriptsize\overrightarrow { -  - } }}} \right) + $
$\left. {{k_{\scriptsize\overrightarrow { -  + } }}{\alpha _{\scriptsize\overrightarrow { -  + } }}\left( {{\alpha _{\scriptsize\overrightarrow { +  - } }} - {\alpha _{\scriptsize\overrightarrow { +  + } }}} \right)} \right]{x_{\scriptsize +}}{x_{\scriptsize -}} + {k_{\scriptsize\overrightarrow { -  + } }}{\alpha _{\scriptsize\overrightarrow { -  + } }}{\beta _{\scriptsize\overrightarrow { +  - } }}x_{\scriptsize -}^2\!$ and $\!B\left( {{x_{\scriptsize +}}} \right) \!$
$ = {k_{\scriptsize\overrightarrow { +  - } }}{\alpha _{\scriptsize\overrightarrow { +  - } }}{\beta _{\scriptsize\overrightarrow { -  + } }}x_{\scriptsize +}^2 + \left[ {{\beta _{\scriptsize\overrightarrow { -  + } }}\left( {{k_{\scriptsize\overrightarrow { +  + } }}{\alpha _{\scriptsize\overrightarrow { +  + } }} + {k_{\scriptsize\overrightarrow { +  - } }}{\alpha _{\scriptsize\overrightarrow { -  - } }}} \right)} \right.+ {k_{\scriptsize\overrightarrow { +  - } }}{\alpha _{\scriptsize\overrightarrow { +  - } }}$
$\left. \left( {{\alpha _{\scriptsize\overrightarrow { -  + } }} - {\alpha _{\scriptsize\overrightarrow { -  - } }}} \right) \right]{x_{\scriptsize +}}{x_{\scriptsize -}} + {k_{\scriptsize\overrightarrow { +  + } }}{\alpha _{\scriptsize\overrightarrow { -  + } }}{\beta _{\scriptsize\overrightarrow { -  - } }}x_{\scriptsize -}^2$ are positive, provided that $\forall {\alpha _{\scriptsize\overrightarrow {XY} }} $, ${\beta _{\scriptsize\overrightarrow {XY} }}$, ${k_{\scriptsize\overrightarrow {XY} }} \in \left( {0,1} \right),\overrightarrow {XY}  \in S$ (See Supplemental Material for more details). It implies that the opinions are driven by the probability of breaking directed links ${k_{\scriptsize\overrightarrow {XY} }}$ and the probability of choosing the student to reconnect ${\alpha _{\scriptsize\overrightarrow {XY} }}$. Multiplying $C\left( x_{\scriptsize +} \right) = A\left( {{x_{\scriptsize +}}} \right)B\left( {{x_{\scriptsize +}}} \right)k_{\scriptsize\overrightarrow { +  + } }^{ - 1}k_{\scriptsize\overrightarrow { -  - } }^{ - 1} $ which is positive
on the right side does not alter the asymptotic dynamics, i.e., the fixed point and its stability. We end up with the equation
\begin{equation}
\label{eq.3}
\begin{array}{l}
{{\dot x}_ {\scriptsize +} } = {x_{\scriptsize +}}{x_{\scriptsize -}}\left[ {\left( {{u_1}x_{\scriptsize +}^3 + {u_2}x_{\scriptsize +}^2{x_{\scriptsize -}} + {u_3}{x_{\scriptsize +}}x_{\scriptsize -}^2 + {u_4}x_{\scriptsize -}^3} \right)} \right.\\
\quad \ \left. { - \left( {{v_1}x_{\scriptsize +}^3 + {v_2}x_{\scriptsize +}^2{x_{\scriptsize -}} + {v_3}{x_{\scriptsize +}}x_{\scriptsize -}^2 + {v_4}x_{\scriptsize -}^3} \right)} \right],
\end{array}
\end{equation}
where ${u_1} = \left( {{{{k_{\scriptsize\overrightarrow { +  - } }}} \mathord{\left/
			{\vphantom {{{k_{\scriptsize\overrightarrow { +  - } }}} {{k_{\scriptsize\overrightarrow { +  + } }}}}} \right.
			\kern-\nulldelimiterspace} {{k_{\scriptsize\overrightarrow { +  + } }}}}} \right)\alpha _{\tiny\overrightarrow { +  - } }^2{\beta _{\scriptsize\overrightarrow { +  + } }}{\beta _{\scriptsize\overrightarrow { -  + } }}, {u_2} = {\alpha _{\scriptsize\overrightarrow { +  + } }}{\alpha _{\scriptsize\overrightarrow { +  - } }}{\beta _{\scriptsize\overrightarrow { +  + } }}{\beta _{\scriptsize\overrightarrow { -  + } }} + \left( {{{{k_{\scriptsize\overrightarrow { +  - } }}} \mathord{\left/
		{\vphantom {{{k_{\scriptsize\overrightarrow { +  - } }}} {{k_{\scriptsize\overrightarrow { +  + } }}}}} \right.
		\kern-\nulldelimiterspace} {{k_{\scriptsize\overrightarrow { +  + } }}}}} \right)\left[ {{\alpha _{\scriptsize\overrightarrow { +  - } }}{\alpha _{\scriptsize\overrightarrow { -  - } }}{\beta _{\scriptsize\overrightarrow { +  - } }}} \right.\left( {{\beta _{\scriptsize\overrightarrow { +  + } }} + } \right.$\\$\!\!\!\left. {\left. {{\beta _{\scriptsize\overrightarrow { -  + } }}} \right) + {\alpha _{\scriptsize\overrightarrow { +  - } }}{\alpha _{\scriptsize\overrightarrow { -  + } }}{\beta _{\scriptsize\overrightarrow { +  + } }}\left( {{\alpha _{\scriptsize\overrightarrow { +  - } }} - {\alpha _{\scriptsize\overrightarrow { -  - } }}} \right)} \right], {u_3} \!=\! {\alpha _{\scriptsize\overrightarrow { +  + } }}{\alpha _{\scriptsize\overrightarrow { -  - } }}{\beta _{\scriptsize\overrightarrow { +  - } }}{\beta _{\scriptsize\overrightarrow { -  + } }} + {\alpha _{\scriptsize\overrightarrow { +  - } }}{\alpha _{\scriptsize\overrightarrow { -  + } }}{\beta _{\scriptsize\overrightarrow { +  + } }}{\beta _{\scriptsize\overrightarrow { -  - } }} + \left( {{{{k_{\scriptsize\overrightarrow { +  - } }}} \mathord{\left/
			{\vphantom {{{k_{\scriptsize\overrightarrow { +  - } }}} {{k_{\scriptsize\overrightarrow { +  + } }}}}} \right.
			\kern-\nulldelimiterspace} {{k_{\scriptsize\overrightarrow { +  + } }}}}} \right)$\\$\left[ {\alpha _{\tiny\overrightarrow { -  - } }^2\beta _{\tiny\overrightarrow { +  - } }^2 + {\alpha _{\scriptsize\overrightarrow { -  + } }}{\alpha _{\scriptsize\overrightarrow { -  - } }}{\beta _{\scriptsize\overrightarrow { +  - } }}\left( {{\alpha _{\scriptsize\overrightarrow { +  - } }} - {\alpha _{\scriptsize\overrightarrow { -  - } }}} \right)} \right], {u_4} = {\alpha _{\scriptsize\overrightarrow { -  + } }}{\alpha _{\scriptsize\overrightarrow { -  - } }}{\beta _{\scriptsize\overrightarrow { +  - } }}{\beta _{\scriptsize\overrightarrow { -  - } }}$ and ${v_1}={\alpha _{\scriptsize\overrightarrow { +  + } }}{\alpha _{\scriptsize\overrightarrow { +  - } }}{\beta _{\scriptsize\overrightarrow { +  + } }}{\beta _{\scriptsize\overrightarrow { -  + } }}$, ${v_2} = {{\alpha _{\scriptsize\overrightarrow { +  + } }}{\alpha _{\scriptsize\overrightarrow { -  - } }}}$ ${\beta _{\scriptsize\overrightarrow { +  - } }}{\beta _{\scriptsize\overrightarrow { -  + } }} + {\alpha _{\scriptsize\overrightarrow { +  - } }}{\alpha _{\scriptsize\overrightarrow { -  + } }}{\beta _{\scriptsize\overrightarrow { +  + } }}{\beta _{\scriptsize\overrightarrow { -  - } }} + \left( {{{{k_{\scriptsize\overrightarrow { -  + } }}} \mathord{\left/
			{\vphantom {{{k_{\scriptsize\overrightarrow { -  + } }}} {{k_{\scriptsize\overrightarrow { -  - } }}}}} \right.
			\kern-\nulldelimiterspace} {{k_{\scriptsize\overrightarrow { -  - } }}}}} \right)\left[ {\alpha _{\tiny\overrightarrow { +  + } }^2\beta _{\tiny\overrightarrow { -  + } }^2 + } \right.{{\alpha _{\scriptsize\overrightarrow { +  + } }}}\left. {{\alpha _{\scriptsize\overrightarrow { +  - } }}{\beta _{\scriptsize\overrightarrow { -  + } }}\left( {{\alpha _{\scriptsize\overrightarrow { -  + } }} - {\alpha _{\scriptsize\overrightarrow { +  + } }}} \right)} \right]$, ${v_3} = {\alpha _{\scriptsize\overrightarrow { -  + } }}{\alpha _{\scriptsize\overrightarrow { -  - } }}{\beta _{\scriptsize\overrightarrow { +  - } }}{\beta _{\scriptsize\overrightarrow { -  - } }} + \left( {{{{k_{\scriptsize\overrightarrow { -  + } }}} \mathord{\left/
			{\vphantom {{{k_{\scriptsize\overrightarrow { -  + } }}} {{k_{\scriptsize\overrightarrow { -  - } }}}}} \right.
			\kern-\nulldelimiterspace} {{k_{\scriptsize\overrightarrow { -  - } }}}}} \right){\left[ {{\alpha _{\scriptsize\overrightarrow { +  + } }}{\alpha _{\scriptsize\overrightarrow { -  + } }}{\beta _{\scriptsize\overrightarrow { -  + } }}\left( {{\beta _{\scriptsize\overrightarrow { +  - } }} + {\beta _{\scriptsize\overrightarrow { -  - } }}} \right) + {\alpha _{\scriptsize\overrightarrow { +  - } }}{\alpha _{\scriptsize\overrightarrow { -  + } }}{\beta _{\scriptsize\overrightarrow { -  - } }}\left( {{\alpha _{\scriptsize\overrightarrow { -  + } }} - {\alpha _{\scriptsize\overrightarrow { +  + } }}} \right)} \right]},$ ${v_4} = \left( {{{{k_{\scriptsize\overrightarrow { -  + } }}} \mathord{\left/
			{\vphantom {{{k_{\scriptsize\overrightarrow { -  + } }}} {{k_{\scriptsize\overrightarrow { -  - } }}}}} \right.
			\kern-\nulldelimiterspace} {{k_{\scriptsize\overrightarrow { -  - } }}}}} \right)\alpha _{\tiny\overrightarrow { -  + } }^2{\beta _{\scriptsize\overrightarrow { +  - } }}{\beta _{\scriptsize\overrightarrow { -  - } }}$. Eq. \hyperref[eq.3]{(3)} is a replicator equation whose payoff matrix is given by
\begin{table}[H]
	\centering
	\caption{\label{table 1}The fate of opinions is captured by the emergent payoff matrix of four-player two-strategy game. }
		\begin{tabular}{ccccc}
			\br
			Individual(s)&\qquad3+&\qquad2+&\qquad1+&\qquad0+\\
			\br
			%\mbox{Three}&\mbox{Four}&\mbox{Five}\\
			+&\qquad$u_1$&\qquad\mbox{$u_2/3$}&\qquad\mbox{$u_3/3$}&\qquad\mbox{$u_4$}\\
			$-$&\qquad$v_1$&\qquad$v_2/3$&\qquad$v_3/3$&\qquad$v_4$\\
			\br
		\end{tabular}
%	\end{ruledtabular}
\end{table}

Let ${f_ {\scriptsize +} }\left( {{x_{\scriptsize +}}} \right) = {u_1}x_{\scriptsize +}^3 + {u_2}x_{\scriptsize +}^2{x_{\scriptsize -}} + {u_3}{x_{\scriptsize +}}x_{\scriptsize -}^2 + {u_4}x_{\scriptsize -}^3$ which refers to the average payoff of opinion $+$ and ${f_ {\scriptsize -} }\left( {{x_{\scriptsize +}}} \right) = {v_1}x_{\scriptsize +}^3 + {v_2}x_{\scriptsize +}^2{x_{\scriptsize -}} + {v_3}{x_{\scriptsize +}}x_{\scriptsize -}^2 + {v_4}x_{\scriptsize -}^3$ which refers to the average payoff of opinion $-$. This implies for large population size, the voting behavior on the directed dynamical network is captured by the replicator equation of a four-player two-strategy game with payoff matrix \hyperref[table 1]{Table 1} in the well-mixed population. For example, the payoff of an individual with opinion $+$ is $u_1$ if the focal individual interacts with three individuals with opinion $+$. There are eight parameters in our model, i.e., ${\alpha_{\scriptsize\overrightarrow {XY} }}$ and ${k_{\scriptsize\overrightarrow {XY} }}$, where $\overrightarrow {XY}  \in S$.

\subsection{Emergent two-player games: predicting bistability and coexistence of opinions}
In the linking dynamics, we have two classes of parameters, i.e., the probability of choosing source nodes ${\alpha_{\scriptsize\overrightarrow {XY} }}$ and the probability of breaking directed links ${k_{\scriptsize\overrightarrow {XY} }}$. We analyze the fate of the opinions with the two classes of parameters, respectively.

\subsubsection{The same probability of choosing source nodes}
We assume that the probabilities of rewiring nodes are equal, i.e., there exists an $\alpha \in \left( {0,1} \right)$ such that ${\alpha_{\scriptsize\overrightarrow {X Y} }} = \alpha $, where $\overrightarrow {XY}  \in S$. Substituting it into \hyperref[table 1]{Table 1}, we obtain
$$\left( {\begin{array}{*{20}{c}}
	{{u_1}}\\
	{{{{u_2}} \mathord{\left/
				{\vphantom {{{u_2}} 3}} \right.
				\kern-\nulldelimiterspace} 3}}\\
	{{{{u_3}} \mathord{\left/
				{\vphantom {{{u_3}} 3}} \right.
				\kern-\nulldelimiterspace} 3}}\\
	{{u_4}}
	\end{array}} \right) = \displaystyle\frac{{{\alpha ^2}{{\left( {1 - \alpha } \right)}^2}}}{3}\left( {\begin{array}{*{20}{c}}
	3&{}&0\\
	2&{}&1\\
	1&{}&2\\
	0&{}&3
	\end{array}} \right) \cdot \left( {\begin{array}{*{20}{c}}
	{\displaystyle\frac{{{k_{\scriptsize\overrightarrow { +  - } }}}}{{{k_{\scriptsize\overrightarrow { +  + } }}}}}\\
	1
	\end{array}} \right)$$
and $$\left( {\begin{array}{*{20}{c}}
	{{v_1}}\\
	{{{{v_2}} \mathord{\left/
				{\vphantom {{{v_2}} 3}} \right.
				\kern-\nulldelimiterspace} 3}}\\
	{{{{v_3}} \mathord{\left/
				{\vphantom {{{v_3}} 3}} \right.
				\kern-\nulldelimiterspace} 3}}\\
	{{v_4}}
	\end{array}} \right) = \displaystyle\frac{{{\alpha ^2}{{\left( {1 - \alpha } \right)}^2}}}{3}\left( {\begin{array}{*{20}{c}}
	3&{}&0\\
	2&{}&1\\
	1&{}&2\\
	0&{}&3
	\end{array}} \right) \cdot \left( {\begin{array}{*{20}{c}}
	1\\
	{\displaystyle\frac{{{k_{\scriptsize\overrightarrow { -  + } }}}}{{{k_{\scriptsize\overrightarrow { -  - } }}}}}
	\end{array}} \right)$$

It implies that, for example, the payoff of one individual $+$ who meets three other individuals with opinion $+$ in the four-player game is equal to sum of the payoff of one individual $+$ who meets one individual $+$ in a two-player game, i.e., ${u_1} = {\alpha ^2}{\left( {1 - \alpha } \right)^2}{{{k_{\scriptsize\overrightarrow { +  - } }}} \mathord{\left/
		{\vphantom {{{k_{\scriptsize\overrightarrow { +  - } }}} {{k_{\scriptsize\overrightarrow { +  + } }}}}} \right.
		\kern-\nulldelimiterspace} {{k_{\scriptsize\overrightarrow { +  + } }}}}$. Therefore, the four-player two-strategy game degenerates to the two-player two-strategy game, whose payoff matrix is

\begin{equation}
\label{eq.4}
\begin{array}{l}
M_{\rm opinion } = \begin{array}{*{20}{c}}
{}&{\begin{array}{*{20}{c}}
	+ &{}&&{} -
	\end{array}}\\
{\begin{array}{*{20}{c}}
	+ \\
	{}\\
	-
	\end{array}}&{\left( {\begin{array}{*{20}{c}}
		{\displaystyle\frac{{k_{\scriptsize\overrightarrow { +  - } }}}{{{k_{\scriptsize\overrightarrow { +  + } }}}}}&1\\
		1&{\displaystyle\frac{{k_{\scriptsize\overrightarrow { -  + } }}}{{{k_{\scriptsize\overrightarrow { -  - } }}}}}
		\end{array}} \right)}
\end{array}.
\end{array}
\end{equation}
The emergent payoff matrix is independent on $\alpha$. Intuitively, the payoff of an individual $+$ against an individual $+$ is proportional to ${{{k_{\scriptsize\overrightarrow { +  - } }}} \mathord{\left/
		{\vphantom {{{k_{\scriptsize\overrightarrow { +  - } }}} {{k_{\scriptsize\overrightarrow { +  + } }}}}} \right.
		\kern-\nulldelimiterspace} {{k_{\scriptsize\overrightarrow { +  + } }}}}$. If ${k_{\scriptsize\overrightarrow { +  - } }}$ is increased solely, then the number of students with opinion $+$ who learn opinion $-$ decreases. A part of these students reconnect to new teachers with opinion $+$ and adopt opinion $+$. Hence the proportion of opinion $+$ increases.

In-group bias is a common phenomenon in the real world, which implies that individuals prefer to interact with those who take the same opinion \cite{1971Social, 1979In, CASTANO2002315}. It can lead to consensus in the population. That is to say, individuals tend to have the same opinion with in-group bias. In our model, in-group bias corresponds to ${k _{\scriptsize\overrightarrow { +  - } }} > {k _{\scriptsize\overrightarrow { +  + } }}$ and $ {k _{\scriptsize\overrightarrow { -  + } }} > {k _{\scriptsize\overrightarrow { -  - } }}$. Students who adopt different opinions from their teachers are more likely to break the directed links than those who adopt the same opinions. The emergent payoff matrix in this case is a coordination game. There is only one internal equilibrium of the replicator equation and it is unstable. Thus all the individuals adopt opinion $+$ if the initial fraction of opinion $+$ exceeds
\begin{equation}
x_ {\rm opinion \kern 1pt \scriptsize +} ^ *  = \displaystyle\frac{{{{{k_{\scriptsize\overrightarrow { -  + } }}} \mathord{\left/
				{\vphantom {{{k_{\scriptsize\overrightarrow { -  + } }}} {{k_{\scriptsize\overrightarrow { -  - } }}}}} \right.
				\kern-\nulldelimiterspace} {{k_{\scriptsize\overrightarrow { -  - } }}}} - 1}}{{{{{k_{\scriptsize\overrightarrow { +  - } }}} \mathord{\left/
				{\vphantom {{{k_{\scriptsize\overrightarrow { +  - } }}} {{k_{\scriptsize\overrightarrow { +  + } }}}}} \right.
				\kern-\nulldelimiterspace} {{k_{\scriptsize\overrightarrow { +  + } }}}}  + {{{k_{\scriptsize\overrightarrow { -  + } }}} \mathord{\left/
				{\vphantom {{{k_{\scriptsize\overrightarrow { -  + } }}} {{k_{\scriptsize\overrightarrow { -  - } }}}}} \right.
				\kern-\nulldelimiterspace} {{k_{\scriptsize\overrightarrow { -  - } }}} -2 }}}.
\end{equation}
Otherwise all, the individuals reach a consensus on opinion $-$. It prevents the homogenization of opinions.

The out-group bias implies that individuals prefer to interact with those who adopt different opinions \cite{1971Social, 1979In, CASTANO2002315}. In a large campaign, it is important that the chiefs focus on how to convert voters from the other camp to their own. Out-group bias in our model refers to ${k _{\scriptsize\overrightarrow { +  - } }} < {k _{\scriptsize\overrightarrow { +  + } }}$ and $ {k _{\scriptsize\overrightarrow { -  + } }} < {k _{\scriptsize\overrightarrow { -  - } }}$. The payoff matrix refers to a coexistence game. Standard analysis shows that there is only one internal stable equilibrium $x_{\rm opinion \kern 1pt \scriptsize +}^ *$ of the replicator equation. In other words, opinion $+$ and opinion $-$ coexist if they coexist in the beginning. The network has many directed links with inconsistent opinions, i.e., $\overrightarrow { +  - } $ and $\overrightarrow { -  + } $. Based on stable regimes, if ${k _{\scriptsize\overrightarrow { +  + } }}$ is decreasing or ${k _{\scriptsize\overrightarrow { -  - } }}$ is increasing, then the final fraction of opinion $+$ increases [Figs. \hyperref[tu3]{3(a)} and \hyperref[tu3]{3(b)}]. Other cases are listed in Supplemental Material. Therefore, if the chiefs with opinion $+$ would like to increase the size of their camp, then it can be achieved by decreasing ${k _{\scriptsize\overrightarrow { +  + } }}$ or increasing ${k _{\scriptsize\overrightarrow { -  - } }}$. That is to say, increasing the number of students on the opinion $+$ or decreasing the number of students on the opposite side.

\subsubsection{The same probability of breaking directed links}
We assume that the probabilities of breaking directed links are equal, i.e., there exists a $k \in \left( {0,1} \right)$ such that ${k_{\scriptsize\overrightarrow {X Y} }} = k $, where $\overrightarrow {XY}  \in S$. It implies that the type of the directed links is not taken into account when the links are broken. Substituting ${k_{\scriptsize\overrightarrow {X Y} }} = k $ into Eq. \hyperref[eq.3]{(3)}, we find ${{\dot x}_ {\scriptsize +} } = D\left( {{x_{\scriptsize +}}} \right){x_{\scriptsize +}}{x_{\scriptsize -}}\left[ {\left( {{\alpha _{\scriptsize\overrightarrow { +  - } }}{\beta _{\scriptsize\overrightarrow { -  + } }}{x_{\scriptsize +}} + } \right.} \right.$
$\left. {\left. {{\alpha _{\scriptsize\overrightarrow { -  - } }}{\beta _{\scriptsize\overrightarrow { +  - } }}{x_{\scriptsize -}}} \right) - \left( {{\alpha _{\scriptsize\overrightarrow { +  + } }}{\beta _{\scriptsize\overrightarrow { -  + } }}{x_{\scriptsize +}} + {\alpha _{\scriptsize\overrightarrow { -  + } }}{\beta _{\scriptsize\overrightarrow { +  - } }}{x_{\scriptsize -}}} \right)} \right]$, where $D\left( {{x_{\scriptsize +}}} \right)$
$ = {\beta _{\scriptsize\overrightarrow { +  + } }}{\alpha _{\scriptsize\overrightarrow { +  - } }}x_{\scriptsize +}^2 + \left( {{\alpha _{\scriptsize\overrightarrow { +  + } }}{\beta _{\scriptsize\overrightarrow { -  + } }} + {\alpha _{\scriptsize\overrightarrow { -  - } }}{\beta _{\scriptsize\overrightarrow { +  - } }}} \right){x_{\scriptsize +}}{x_{\scriptsize -}} + {\alpha _{\scriptsize\overrightarrow { -  + } }}{\beta _{\scriptsize\overrightarrow { -  - } }}x_{\scriptsize -}^2$ is positive. Similarly, we end up with a replicator equation, i.e., ${{\dot x}_ {\scriptsize +} } = {x_{\scriptsize +}}{x_{\scriptsize -}}\left[ {\left( {{\alpha _{\scriptsize\overrightarrow { +  - } }}{\beta _{\scriptsize\overrightarrow { -  + } }}{x_{\scriptsize +}} + } \right.}  {{\alpha _{\scriptsize\overrightarrow { -  - } }}{\beta _{\scriptsize\overrightarrow { +  - } }}{x_{\scriptsize -}}} \right)-
\left( {{\alpha _{\scriptsize\overrightarrow { +  + } }}{\beta _{\scriptsize\overrightarrow { -  + } }}{x_{\scriptsize +}}} {\left. { + {\alpha _{\scriptsize\overrightarrow { -  + } }}{\beta _{\scriptsize\overrightarrow { +  - } }}{x_{\scriptsize -}}} \right)} \right]$, whose payoff matrix is the two-player two-strategy game
\begin{equation}
\label{eq.6}
{\begin{array}{*{20}{c}}
	{}\\
	{R_ {\rm opinion} = }
	\end{array}}
\begin{array}{*{20}{c}}
{}&{\begin{array}{*{20}{c}}
	+ &{}&{}&{}&\quad -
	\end{array}}\\
{\begin{array}{*{20}{c}}
	+ \\
	-
	\end{array}}\!\!\!\!\!\!&{\left( {\begin{array}{*{20}{c}}
		{{\alpha _{\scriptsize\overrightarrow { +  - } }} {\beta _{\scriptsize\overrightarrow { -  + } }} }&{{\alpha _{\scriptsize\overrightarrow { -  - } }} {\beta _{\scriptsize\overrightarrow { +  - } }} }\\
		{{\alpha _{\scriptsize\overrightarrow { +  + } }} {\beta _{\scriptsize\overrightarrow { -  + } }} }&{{\alpha _{\scriptsize\overrightarrow { -  + } }} {\beta _{\scriptsize\overrightarrow { +  - } }} }
		\end{array}} \right)}
\end{array}.
\end{equation}

Noteworthily, the emergent payoff matrix is independent on $k$ and the payoff entry ${R_{XY}}$ is proportional to ${l_{\scriptsize\overrightarrow {YX} }}$, i.e., the number of directed links $\overrightarrow {YX} $. For example, the payoff of an individual $+$ meeting an individual $-$ is proportional to ${l_{\scriptsize\overrightarrow { -  + } }}$, which refers to the number of students $-$ who have teachers with opinion $+$. Here is an intuitive explanation: if ${\alpha _{\scriptsize\overrightarrow { -  - } }}$ increases solely, then a part of students with opinion $-$ reconnect to the new teachers with opinion $+$. Hence ${l_{\scriptsize\overrightarrow { -  + } }}$ increases.

Similarly, we discuss the following two cases. We address a coordination game with ${\alpha _{\scriptsize\overrightarrow { +  - } }} > {\alpha _{\scriptsize\overrightarrow { +  + } }}$ and $ {\alpha _{\scriptsize\overrightarrow { -  + } }} > {\alpha _{\scriptsize\overrightarrow { -  - } }}$. In this scenario, there is an unstable internal equilibrium given by
\begin{equation}
y_{\rm opinion \kern 1pt \scriptsize +}^ *  = \frac{{{\beta _{\scriptsize\overrightarrow { +  - } }}\left( {{\alpha _{\scriptsize\overrightarrow { -  - } }} - {\alpha _{\scriptsize\overrightarrow { -  + } }}} \right)}}{{{\beta _{\scriptsize\overrightarrow { -  + } }}\left( {{\alpha _{\scriptsize\overrightarrow { +  + } }} - {\alpha _{\scriptsize\overrightarrow { +  - } }}} \right) + {\beta _{\scriptsize\overrightarrow { +  - } }}\left( {{\alpha _{\scriptsize\overrightarrow { -  - } }} - {\alpha _{\scriptsize\overrightarrow { -  + } }}} \right)}}
\end{equation}
The individuals reach a consensus with opinion $+$ if the initial fraction of opinion $+$ exceeds $y_{\rm opinion \kern 1pt  \scriptsize +}^ *$. Otherwise, it reaches a consensus with opinion $-$.

We study a coexistence game defined by $\alpha_{\scriptsize\overrightarrow { +  - } } < \alpha_{\scriptsize\overrightarrow { +  + } }$ and $\alpha_{\scriptsize\overrightarrow { -  + } } < \alpha_{\scriptsize\overrightarrow { -  - } }$. In this case, opinion $+$ and opinion $-$ coexist for a long time if they coexist in the beginning. If ${\alpha_{\scriptsize\overrightarrow { +  + } }}$ is decreasing or ${\alpha_{\scriptsize\overrightarrow { -  - } }}$ is increasing, then the fraction fraction of opinion $+$ increases [Figs. \hyperref[tu3]{3(c)} and \hyperref[tu3]{3(d)}]. And other cases see Supplemental Material for details.

\begin{figure}
	\includegraphics[scale=0.32]{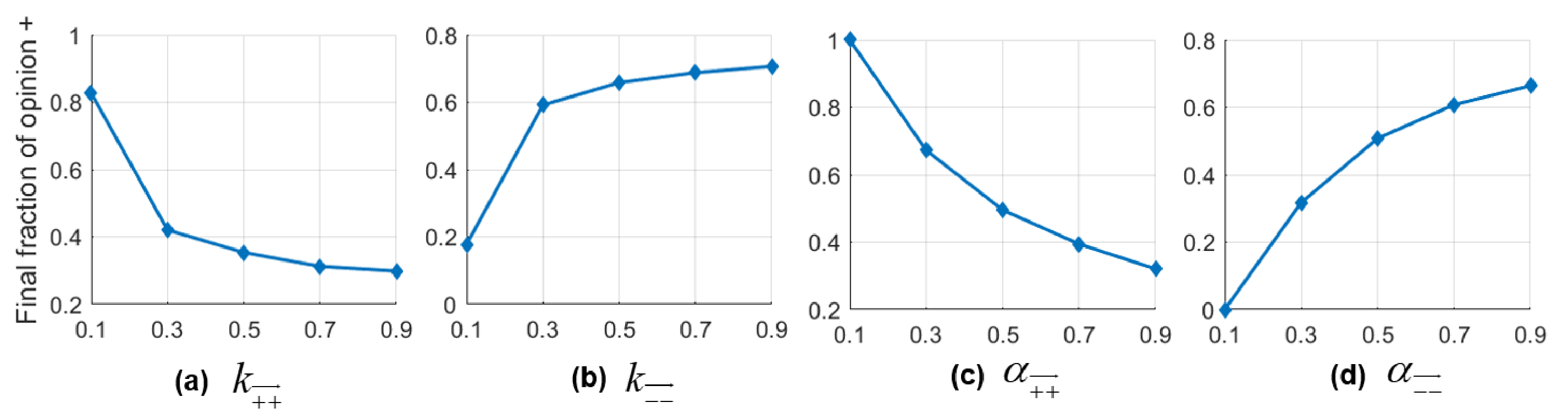}% Here is how to import EPS art
	\caption{\label{tu3}\textbf{The fate of opinions.} We predict the fate of opinion $+$ in the voter model on the directed evolving network. The proportion of one opinion increases when the number of students on the other opinion decreases. That is, the final fraction of opinion $+$ increases as the number of students on opinion $-$ decreases. If $x_{\rm opinion \kern 1pt \scriptsize\protect +}^ *$ is stable, then $x_{\rm opinion \kern 1pt \scriptsize\protect +}^ *$ increases as ${k _{\scriptsize\protect\overrightarrow { +  + } }}$ decreases or ${k _{\scriptsize\protect\overrightarrow { -  - } }}$ increases. If $y_{\rm opinion \kern 1pt \scriptsize\protect +}^ *$ is stable, then $y_{\rm opinion \kern 1pt  \scriptsize\protect +}^ *$ increases as ${\alpha _{\scriptsize\protect\overrightarrow { +  + } }}$ decreases or ${\alpha _{\scriptsize\protect\overrightarrow { -  - } }}$ increases. Parameters: We focus on $x_{\rm opinion \kern 1pt \scriptsize\protect +}^ *$ or $y_{\rm opinion \kern 1pt \scriptsize\protect +}^ *$ is stable internal equilibrium for out-group bias. Hence, we set $k_{\scriptsize\protect\overrightarrow { +  - } } < k_{\scriptsize\protect\overrightarrow { +  + } }$ and $k_{\scriptsize\protect\overrightarrow { -  + } } < k_{\scriptsize\protect\overrightarrow { -  - } }$ when ${\alpha_{\scriptsize\protect\overrightarrow { X  Y } }} = \alpha = 0.5$ or $\alpha_{\scriptsize\protect\overrightarrow { +  - } } < \alpha_{\scriptsize\protect\overrightarrow { +  + } }$ and $\alpha_{\scriptsize\protect\overrightarrow { -  + } } < \alpha_{\scriptsize\protect\overrightarrow { -  - } }$ when ${k_{\scriptsize\protect\overrightarrow { X  Y } }} = k = 0.5$. (a) ${\alpha_{\scriptsize\protect\overrightarrow { X  Y } }} = \alpha = 0.5,$ ${k _{\scriptsize\protect\overrightarrow { +  - } }} = 0.1, {k _{\scriptsize\protect\overrightarrow { -  + } }} = 0.3$ and ${k _{\scriptsize\protect\overrightarrow { -  - } }} = 0.6$. (b) ${\alpha_{\scriptsize\protect\overrightarrow { X  Y } }} = \alpha = 0.5,$ ${k _{\scriptsize\protect\overrightarrow { +  + } }} = 0.6, {k _{\scriptsize\protect\overrightarrow { +  - } }} = 0.3$ and ${k _{\scriptsize\protect\overrightarrow { -  + } }} = 0.1$. (c) ${k_{\scriptsize\protect\overrightarrow { X  Y } }} = k = 0.5,$ ${\alpha _{\scriptsize\protect\overrightarrow { +  - } }} = 0.1, {\alpha _{\scriptsize\protect\overrightarrow { -  + } }} = 0.3$ and ${\alpha _{\scriptsize\protect\overrightarrow { -  - } }} = 0.6$. (d) ${k_{\scriptsize\protect\overrightarrow { X  Y } }} = k = 0.5,$ ${\alpha _{\scriptsize\protect\overrightarrow { +  + } }} = 0.6, {\alpha _{\scriptsize\protect\overrightarrow { +  - } }} = 0.3$ and ${\alpha _{\scriptsize\protect\overrightarrow { -  + } }} = 0.1$. We run ${10^6}$ rounds of the simulation and set the millionth result as the final fraction of opinion $+$. The initial state is $x_ {\scriptsize +} = 0.5$. For each data point, it is averaged over 100 independent runs. We set $N = 100, L = 4$ and $w = 0.01$. }
\end{figure}

\subsection{Emergent multi-player games: complexity analysis}
In subsection A, the four-player two-strategy game degenerates to the two-player two-strategy game provided that there are $\alpha  \in \left( {0,1} \right)$ and $k \in \left( {0,1} \right)$ such that ${\alpha_{\scriptsize\overrightarrow {X Y} }} =  \alpha $ or ${k_{\scriptsize\overrightarrow {X Y} }} = k $ for $\forall \overrightarrow {XY}  \in S$. But what is the complexity of our model? If ${u_1}>{v_1}$, ${u_2}<{v_2}$, ${u_3}>{v_3}$ and ${u_4}<{v_4}$ (or ${u_1}<{v_1}$, ${u_2}>{v_2}$, ${u_3}<{v_3}$ and ${u_4}>{v_4}$) are satisfied in \hyperref[table 1]{Table 1}, ${f_ + }\left( {{x_ {\scriptsize +} }} \right) - {f_ - }\left( {{x_ {\scriptsize +} }} \right)$ changes the sign three times with respect to ${x_ {\scriptsize +} }$ when non-zero coefficients are arranged from highest to lowest according to the power of ${x_ {\scriptsize +} }$. Based on Descartes’ rule of signs \cite{2008Polynomials}, there are one or three roots, i.e., one internal equilibrium or three internal equilibria. We choose one parameter at random from ${\alpha_{\scriptsize\overrightarrow {X Y} }}$ and ${k_{\scriptsize\overrightarrow {X Y} }}$ respectively and make them equal. And we keep the other six parameters equal. We prove that it does not satisfy the condition of changing the sign three times (See Supplemental Material for details). Thus, to reveal the complexity, more parameters are needed to be unequal.

We find a set of parameters, i.e., ${k_{\scriptsize\overrightarrow { +  + } }} = \rho ,{k_{\scriptsize\overrightarrow { +  - } }} = \rho ,{k_{\scriptsize\overrightarrow { -  + } }} = \rho/4,{k_{\scriptsize\overrightarrow { -  - } }} = \rho ,{\alpha _{\scriptsize\overrightarrow { +  + } }} = \rho/2,{\alpha _{\scriptsize\overrightarrow { +  - } }} = \rho ,{\alpha _{\scriptsize\overrightarrow { -  + } }} = 2\rho$ and ${\alpha _{\scriptsize\overrightarrow { -  - } }} = \rho/4$, where $0 < \rho  < 0.5$. These eight parameters are only up to $\rho $. There are three internal equilibria under the condition $\left( {21 - \sqrt {249} } \right)/32 \approx 0.1631 < \rho  < 0.5$, where \bm{${u_1}>{v_1}$}, \bm{${u_2}<{v_2}$}, \bm{${u_3}>{v_3}$} and \bm{${u_4}<{v_4}$}. For example, substituting $\rho  = 0.4$ into \hyperref[table 1]{Table 1}, we obtain

\begin{table}[H]
	\centering
	\caption{The value of payoff matrix.} \label{table 2}
%	\begin{ruledtabular}
		\begin{tabular}{ccccc}
			\br
			Individual(s)&\qquad3+&\qquad2+&\qquad1+&\qquad0+\\
			%\mbox{Three}&\mbox{Four}&\mbox{Five}\\
			\br
			+&\qquad0.0256&\qquad\mbox{0.0379}&\qquad\mbox{0.0836}&\qquad\mbox{0.0432}\\
			$-$&\qquad0.0128&\qquad0.0785&\qquad0.0328&\qquad0.0864\\
			\br
		\end{tabular}
%	\end{ruledtabular}
\end{table}

\begin{figure}[H]
	\centering
	\includegraphics[scale=0.23]{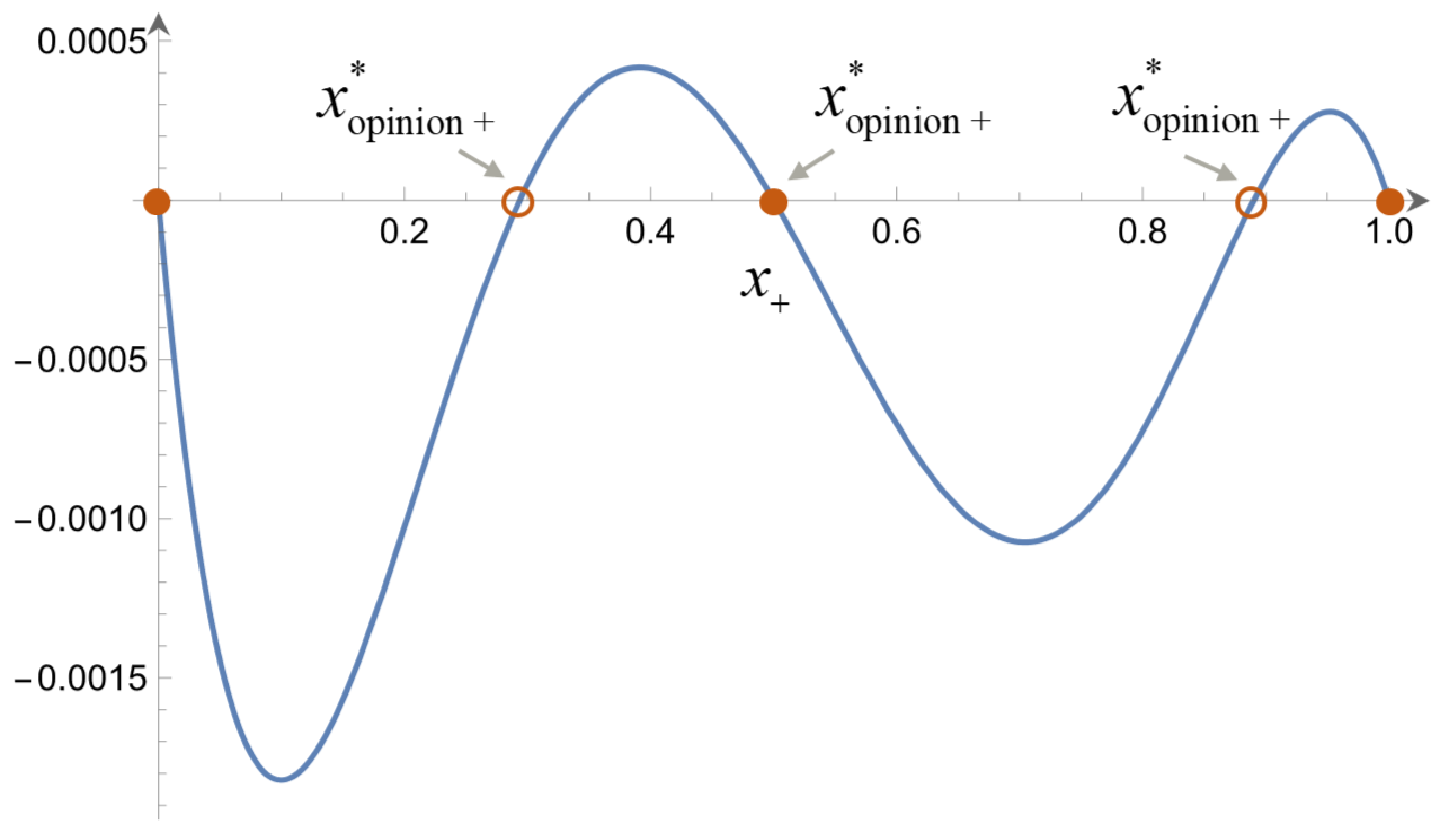}% Here is how to import EPS art
	\caption{\label{tu4}\textbf{Complexity dynamics analysis of opinions.} We have shown that the opinion evolves as a replicator equation of a four-player two-strategy game \hyperref[table 1]{Table 1}. For such a game, there can be at most three internal equilibria. This game is related to the probabilities of choosing nodes and breaking directed links. We can obtain the complexity of opinion dynamics via an evolutionary game approach. There are three internal equilibria, i.e., $x_ {\rm opinion \kern 1pt \scriptsize +} ^ * = 0.29$, $0.5$ and $0.89$ under the condition of $\rho  = 0.4$. $x_ {\rm opinion \kern 1pt \scriptsize +} ^ * = 0.5$ is the only internal stable equilibrium. If the initial fraction of opinion $+$ is about $0.29$, then the individuals reach a consensus on opinion $-$ finally. If the initial fraction of opinion $+$ is between $0.29$ and $0.89$, then the two types of opinions coexist and each one is equally divided. Otherwise, the individuals reach a consensus on opinion $+$. Individuals can maintain diverse opinions or reach a consensus for this game. }
\end{figure}

This four-player two-strategy game has three internal equilibria, i.e., $x_ {\rm opinion \kern 1pt \scriptsize +} ^ * = 0.29$, $0.5$ and $0.89$, as shown in Fig. \hyperref[tu4]{4}. And $x_ {\rm opinion \kern 1pt \scriptsize +} ^ * = 0.5$ is only one internal stable equilibrium. In this case, the final opinions in the population are either diverse or reached a consensus among individuals. Therefore, the complexity of the voter model on the directed evolving network is captured by the four-player two-strategy game.

\subsection{Robustness}
We exchange the direction of learning in the network. For example, if node $B$ points to node $A$, it implies that $A$ unilaterally learns from $B$ and $B$ does not learn from $A$, that is, the target node learns the source node. Therefore, the transition probability that $T_{{x_ {\scriptsize +}}}^ +  = {{{x_{\scriptsize -}}{\pi _{\scriptsize\overrightarrow { +  - } }}} \mathord{\left/
		{\vphantom {{{x_{\scriptsize -}}{\pi _{\scriptsize\overrightarrow { +  - } }}} {\left( {{\pi _{\scriptsize\overrightarrow { +  - } }} + {\pi _{\scriptsize\overrightarrow { -  - } }}} \right)}}} \right.
		\kern-\nulldelimiterspace} {\left( {{\pi _{\scriptsize\overrightarrow { +  - } }} + {\pi _{\scriptsize\overrightarrow { -  - } }}} \right)}}$. The transition probability that $x_{\scriptsize +}$ decreases by $1 / N$ is $T_{{x_ {\scriptsize +}}}^ -  = {{{x_ {\scriptsize +}}{\pi _{\scriptsize\overrightarrow { -  + } }}} \mathord{\left/
		{\vphantom {{{x_{\scriptsize -}}{\pi _{\scriptsize\overrightarrow { -  + } }}} {\left( {{\pi _{\scriptsize\overrightarrow { +  + } }} + {\pi _{\scriptsize\overrightarrow { -  + } }}} \right)}}} \right.
		\kern-\nulldelimiterspace} {\left( {{\pi _{\scriptsize\overrightarrow { +  + } }} + {\pi _{\scriptsize\overrightarrow { -  + } }}} \right)}}$. In this case, we obtain some dual results. Similarly, the voting behavior on the evolving directed network is captured by a four-player two-strategy game whose payoff matrix is given by \hyperref[table 3]{Table 3}, where ${u_1'} \!=\! \left( {{{{k_{\scriptsize\overrightarrow { -  + } }}} \mathord{\left/
			{\vphantom {{{k_{\scriptsize\overrightarrow { -  + } }}} {{k_{\scriptsize\overrightarrow { +  + } }}}}} \right.
			\kern-\nulldelimiterspace} {{k_{\scriptsize\overrightarrow { +  + } }}}}} \right)\alpha _{\scriptsize\overrightarrow { +  + } }{\alpha _{\scriptsize\overrightarrow { +  - } }}{\beta _{\scriptsize\overrightarrow { -  + } }^2}, \!{u_2'} = {\alpha _{\scriptsize\overrightarrow { +  + } }}{\alpha _{\scriptsize\overrightarrow { +  - } }}{\beta _{\scriptsize\overrightarrow { +  + } }}{\beta _{\scriptsize\overrightarrow { -  + } }} +\left( {{{{k_{\scriptsize\overrightarrow { -  + } }}} \mathord{\left/
			{\vphantom {{{k_{\scriptsize\overrightarrow { -  + } }}} {{k_{\scriptsize\overrightarrow { +  + } }}}}} \right.
			\kern-\nulldelimiterspace} {{k_{\scriptsize\overrightarrow { +  + } }}}}} \right) \left[ {\alpha _{\scriptsize\overrightarrow { +  - } }}{\alpha _{\scriptsize\overrightarrow { -  + } }}{\beta _{\scriptsize\overrightarrow { -  + } }} {{\beta _{\scriptsize\overrightarrow { -  - } }} + \alpha _{\scriptsize\overrightarrow { +  + } } \alpha _{\scriptsize\overrightarrow { -  - } }}{\beta _{\scriptsize\overrightarrow { +  - } } \beta _{\scriptsize\overrightarrow { -  + } }} + \alpha _{\scriptsize\overrightarrow { +  + } } \alpha _{\scriptsize\overrightarrow { -  + } } \beta _{\scriptsize\overrightarrow { -  + } } \left( \alpha _{\scriptsize\overrightarrow { +  - } }  - {\alpha _{\scriptsize\overrightarrow { -  - } }} \right) \right], {u_3'} = {\alpha _{\scriptsize\overrightarrow { +  - } }}$\\${\alpha _{\scriptsize\overrightarrow { -  + } }}{\beta _{\scriptsize\overrightarrow { +  + } }}{\beta _{\scriptsize\overrightarrow { -  - } }} + {\alpha _{\scriptsize\overrightarrow { +  + } }}{\alpha _{\scriptsize\overrightarrow { -  - } }} \beta _{\scriptsize\overrightarrow { +  - } } \beta _{\scriptsize\overrightarrow { -  + } }+ \left( {{{{k_{\scriptsize\overrightarrow { -  + } }}} \mathord{\left/
			{\vphantom {{{k_{\scriptsize\overrightarrow { -  + } }}} {{k_{\scriptsize\overrightarrow { +  + } }}}}} \right.
			\kern-\nulldelimiterspace} {{k_{\scriptsize\overrightarrow { +  + } }}}}} \right)\left[ {\alpha _{\scriptsize\overrightarrow { -  + } } \alpha _{\scriptsize\overrightarrow { -  - } } \beta _{\scriptsize\overrightarrow { +  - } } \beta _{\scriptsize\overrightarrow { -  - } } + {\alpha _{\tiny\overrightarrow { -  + } }^{}}{\beta _{\scriptsize\overrightarrow { -  - } }}\left( {{\alpha _{\scriptsize\overrightarrow { +  - } }} - {\alpha _{\scriptsize\overrightarrow { -  - } }}} \right)} \right], $\\$\,{u_4'} = {\alpha _{\scriptsize\overrightarrow { -  + } }}{\alpha _{\scriptsize\overrightarrow { -  - } }}{\beta _{\scriptsize\overrightarrow { +  - } }}{\beta _{\scriptsize\overrightarrow { -  - } }}$ and $\;{v_1'} = {\alpha _{\scriptsize\overrightarrow { +  + } }}{\alpha _{\scriptsize\overrightarrow { +  - } }}{\beta _{\scriptsize\overrightarrow { +  + } }}{\beta _{\scriptsize\overrightarrow { -  + } }}, {v_2'} = {{\alpha _{\scriptsize\overrightarrow { +  + } }}{\alpha _{\scriptsize\overrightarrow { -  - } }}}{\beta _{\scriptsize\overrightarrow { +  - } }}{\beta _{\scriptsize\overrightarrow { -  + } }} + {\alpha _{\scriptsize\overrightarrow { +  - } }}{\alpha _{\scriptsize\overrightarrow { -  + } }}{\beta _{\scriptsize\overrightarrow { +  + } }}$\\${\beta _{\scriptsize\overrightarrow { -  - } }} + \left( {{{{k_{\scriptsize\overrightarrow { +  - } }}} \mathord{\left/
			{\vphantom {{{k_{\scriptsize\overrightarrow { +  - } }}} {{k_{\scriptsize\overrightarrow { -  - } }}}}} \right.
			\kern-\nulldelimiterspace} {{k_{\scriptsize\overrightarrow { -  - } }}}}} \right)\,\left[ \alpha _{\scriptsize\overrightarrow { +  + } } \alpha _{\scriptsize\overrightarrow { +  - } } \right. \beta _{\scriptsize\overrightarrow { +  + } } \beta _{\scriptsize\overrightarrow { -  + } } + {{\alpha _{\scriptsize\overrightarrow { +  - } }^2}} {{\beta _{\scriptsize\overrightarrow { +  + } }}\left( {{\alpha _{\scriptsize\overrightarrow { -  + } }} - {\alpha _{\scriptsize\overrightarrow { +  + } }}} \right)}]\,, {v_3'} \,=\, {\alpha _{\scriptsize\overrightarrow { -  + } }}{\alpha _{\scriptsize\overrightarrow { -  - } }}{\beta _{\scriptsize\overrightarrow { +  - } }}{\beta _{\scriptsize\overrightarrow { -  - } }} + \left( {{{{k_{\scriptsize\overrightarrow { +  - } }}} \mathord{\left/
			{\vphantom {{{k_{\scriptsize\overrightarrow { +  - } }}} {{k_{\scriptsize\overrightarrow { -  - } }}}}} \right.
			\kern-\nulldelimiterspace} {{k_{\scriptsize\overrightarrow { -  - } }}}}} \right) {\left[ {{\alpha _{\scriptsize\overrightarrow { +  - } }}{\alpha _{\scriptsize\overrightarrow { -  + } }}{\beta _{\scriptsize\overrightarrow { +  + } }} {\beta _{\scriptsize\overrightarrow { +  - } }} \,+\, \alpha _{\scriptsize\overrightarrow { +  + } } \alpha _{\scriptsize\overrightarrow { -  - } } \beta _{\scriptsize\overrightarrow { +  - } }{\beta _{\scriptsize\overrightarrow { -  + } }}  \,+\, {\alpha _{\scriptsize\overrightarrow { +  - } }}{\alpha _{\scriptsize\overrightarrow { -  - } }}} \right. }\left. {{\beta _{\scriptsize\overrightarrow { +  - } }}\left( {{\alpha _{\scriptsize\overrightarrow { -  + } }} \,-\, {\alpha _{\scriptsize\overrightarrow { +  + } }}} \right)} \right], {v_4'} = $\\$\left( {{{{k_{\scriptsize\overrightarrow { +  - } }}} \mathord{\left/
			{\vphantom {{{k_{\scriptsize\overrightarrow { +  - } }}} {{k_{\scriptsize\overrightarrow { -  - } }}}}} \right.
			\kern-\nulldelimiterspace} {{k_{\scriptsize\overrightarrow { -  - } }}}}} \right)\alpha _{\scriptsize\overrightarrow { -  + } } \alpha _{\scriptsize\overrightarrow { -  - } } {\beta _{\scriptsize\overrightarrow { +  - } }^2}$.
\begin{table}[H]
	\centering
	\caption{\label{table 3}The payoff matrix of the emergent two-strategy four-player game.}
%	\begin{ruledtabular}
		\begin{tabular}{ccccc}
			\br
			Individual(s)&\qquad3+&\qquad2+&\qquad1+&\qquad0+\\
			%\mbox{Three}&\mbox{Four}&\mbox{Five}\\
			\br
			+&\qquad$u'_1$&\qquad\mbox{$u'_2/3$}&\qquad\mbox{$u'_3/3$}&\qquad\mbox{$u'_4$}\\
			$-$&\qquad$v'_1$&\qquad$v'_2/3$&\qquad$v'_3/3$&\qquad$v'_4$\\
			\br
		\end{tabular}
%	\end{ruledtabular}
\end{table}

The four-player two-strategy game degenerates to the two-player two-strategy game if ${\alpha_{\scriptsize\overrightarrow {X Y} }} = \alpha $, where $\overrightarrow {XY}  \in S$ and $0<\alpha<1$. The payoff matrix is
\begin{equation}
M_{\rm opinion\_dual } = \begin{array}{*{20}{c}}
{}&{\begin{array}{*{20}{c}}
	+ &{}&&{} -
	\end{array}}\\
{\begin{array}{*{20}{c}}
	+ \\
	{}\\
	-
	\end{array}}&{\left( {\begin{array}{*{20}{c}}
		{\displaystyle\frac{{k_{\scriptsize\overrightarrow { -  + } }}}{{{k_{\scriptsize\overrightarrow { +  + } }}}}}&1\\
		1&{\displaystyle\frac{{k_{\scriptsize\overrightarrow { +  - } }}}{{{k_{\scriptsize\overrightarrow { -  - } }}}}}
		\end{array}} \right)}.
\end{array}
\end{equation}
And if ${k_{\scriptsize\overrightarrow {X Y} }} = k $, where $\overrightarrow {XY}  \in S$ and $0<k<1$, the payoff matrix is
\begin{equation}
{\begin{array}{*{20}{c}}
	{}\\
	{R_{\rm opinion\_dual } = }\;\,
	\end{array}}{}
\begin{array}{*{20}{c}}
{}&{\begin{array}{*{20}{c}}
	+ &{}&{}&{}&\quad -
	\end{array}}\\
{\begin{array}{*{20}{c}}\!\!\!\!\!\!
	+ \\
	\!\!\!\!\!\!-
	\end{array}}\!\!\!\!\!\!&{\left( {\begin{array}{*{20}{c}}
		{{\alpha _{\scriptsize\overrightarrow { +  + } }} {\alpha _{\scriptsize\overrightarrow { +  - } }} }&{{\alpha _{\scriptsize\overrightarrow { +  - } }} {\alpha _{\scriptsize\overrightarrow { -  + } }} }\\
		{{\alpha _{\scriptsize\overrightarrow { +  - } }} {\alpha _{\scriptsize\overrightarrow { -  + } }} }&{{\alpha _{\scriptsize\overrightarrow { -  + } }} {\alpha _{\scriptsize\overrightarrow { -  - } }} }
		\end{array}} \right)}.
\end{array}
\end{equation}

\section{Emergent games for the transient topology during the opinion formation}
In the preceding section, we focus on the fate of opinions. Here we address the other side of the coin, i.e., the transient property of the evolving networks.

What are the key topology features that pave the way for the successful invasion? In our model, the in-degree of an individual is equal to its student size, and the out-degree is equal to its teacher size. In the voter model, teachers preach their opinions and students adopt the popular opinions. The in-degree, i.e., the teacher's student size is crucial for spreading the teacher's opinions. Hence, we concentrate on the in-degree.

\begin{figure*}
	~
	\includegraphics[scale=0.28]{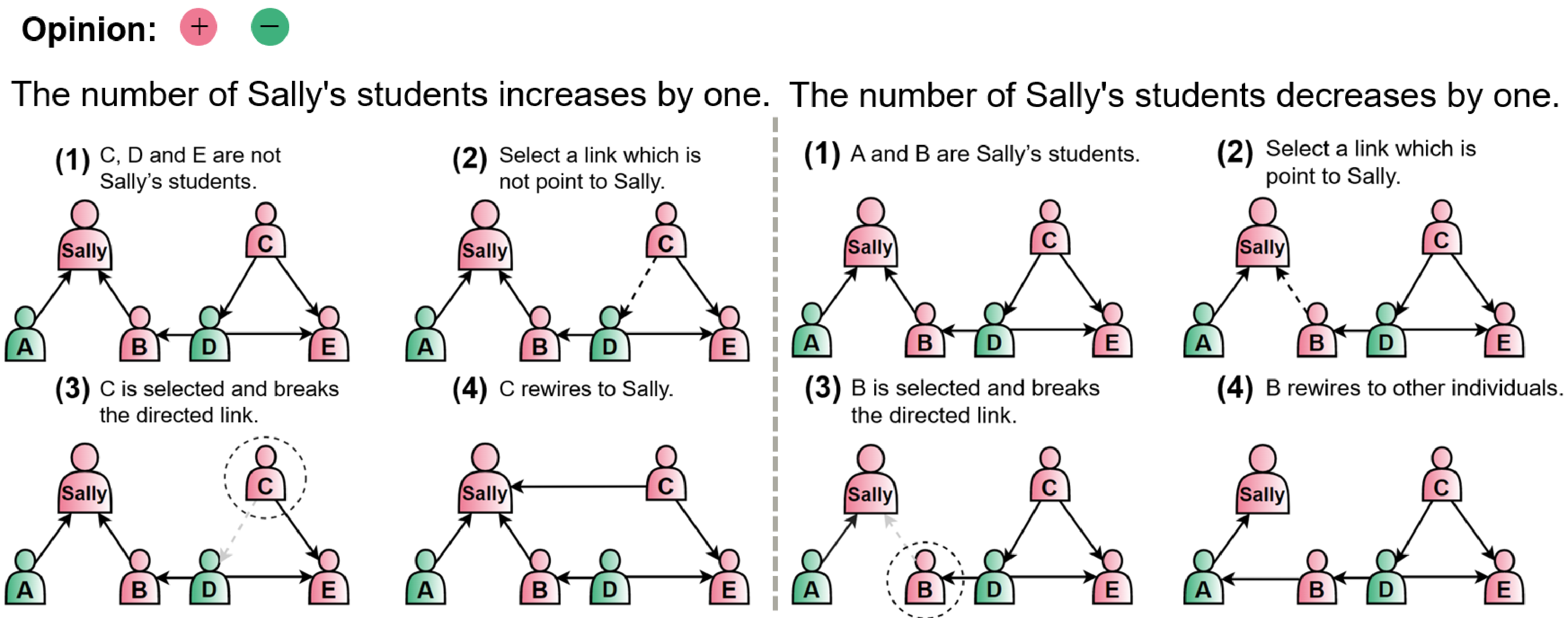}% Here is how to import EPS art
	\caption{\label{tu5}\textbf{Markov transitions of Sally's student size.} \textbf{Left Panel:} The number of Sally's students increases by one. (1) Only part of the directed network is shown here. (2) Select a link $\protect\overrightarrow {XY} $ which is not point to Sally with probability ${{\left( {NL - {d_ {{\rm{in}}\,\scriptsize +} }} \right)} \mathord{\left/
				{\vphantom {{\left( {NL - {d_ {{\rm{in}}\,\scriptsize +} }} \right)} {NL}}} \right.
				\kern-\nulldelimiterspace} {NL}}$. (3) The probability that the type of the selected link is $\protect\overrightarrow {XY} $ depends on ${\pi _S}$. And the probability of selecting the student $X$ is ${\alpha _{\scriptsize\protect\overrightarrow {XY} }}$. Then the student chooses to change the teacher and breaks the link with probability ${k _{\scriptsize\protect\overrightarrow {XY} }}$. (4) Finally, $X$ is connected to Sally with probability ${1 \mathord{\left/
				{\vphantom {1 {\left( {N - 1} \right)}}} \right.
				\kern-\nulldelimiterspace} {\left( {N - 1} \right)}}$. Hence the number of Sally's students increases by one. Without loss of generality, $C$, $D$ and $E$ are not Sally's students. We assume that the directed link $\protect\overrightarrow {CD} $ is selected. The type of $\protect\overrightarrow {CD} $ is $\protect\overrightarrow { +  - } $. Then $C$ is selected and breaks the link with probability ${\alpha _{\scriptsize\protect\overrightarrow { +  - } }}{k _{\scriptsize\protect\overrightarrow { +  - } }}$. Eventually, student $C$ chooses the new teacher Sally. \textbf{Right Panel:} The number of Sally's students decreases by one. (1) Only part of the directed network is shown here. (2) Select a link $\protect\overrightarrow {XY} $ which is point to Sally with probability ${d_ {{\rm{in}}\,\scriptsize +} }/NL$. (3) The probability that the type of the selected link is $\protect\overrightarrow {XY} $ depends on ${\pi _S}$. And the probability of selecting the student $X$ is ${\alpha _{\scriptsize\protect\overrightarrow {XY} }}$. Then the student chooses to change the teacher and breaks the link with probability ${k _{\scriptsize\protect\overrightarrow {XY} }}$. (4) Finally, $X$ is connected to other nodes with probability 1. Hence the number of Sally's students decreases by one. Without loss of generality, $A$ and $B$ are Sally's students. We assume that the directed link $\protect\overrightarrow {BS} $ is selected. The type of $\protect\overrightarrow {BS} $ is $\protect\overrightarrow { +  + } $. Then $B$ is selected and breaks the link with probability ${\alpha _{\scriptsize\protect\overrightarrow { +  + } }}{k _{\scriptsize\protect\overrightarrow { +  + } }}$. Eventually, student $B$ chooses other nodes to find a new teacher $A$.}
\end{figure*}

Suppose there is an individual, named after Sally. Without loss of generality, she adopts the opinion $+$. And she has in-degree ${d_ {{\rm{in}}\,\scriptsize +} }$, i.e., she has ${d_ {{\rm{in}}\,\scriptsize +} }$ students. The in-degree ${d_ {{\rm{in}}\,\scriptsize +} }$ of Sally ranges from $0$ to $N-1$. For our linking dynamics, Sally's in-degree increases or decreases by at most one. If an individual who is not Sally's current student reconnects to her, Sally's in-degree ${d_ {{\rm{in}}\,\scriptsize +} }$ increases by one: firstly, the probability of selecting the directed link $\overrightarrow {XY} $ which is not point to Sally is ${{\left( {NL - {d_ {{\rm{in}}\,\scriptsize +} }} \right)} \mathord{\left/
		{\vphantom {{\left( {NL - {d_ {{\rm{in}}\,\scriptsize +} }} \right)} {NL}}} \right.
		\kern-\nulldelimiterspace} {NL}}$, where $\overrightarrow {XY}  \in S$. Secondly, the stationary distribution of the directed links is ${\pi_S} = \left( {{\pi _{\scriptsize\overrightarrow { +  + } }},{\pi _{\scriptsize\overrightarrow { +  - } }},{\pi _{\scriptsize\overrightarrow { -  + } }},{\pi _{\scriptsize\overrightarrow { -  - } }}} \right)$, which has been given by Eq. \hyperref[eq.A.4]{(A.4)}. Then student $X$ is chosen with probability ${\alpha _{\scriptsize\overrightarrow {XY} }}$ and breaks the directed link $\overrightarrow {XY} $ with probability ${k_{\scriptsize\overrightarrow {XY} }}$. Finally, student $X$ connects to Sally with probability ${1 \mathord{\left/
		{\vphantom {1 {\left( {N - 1} \right)}}} \right.
		\kern-\nulldelimiterspace} {\left( {N - 1} \right)}}$. Thus the transition probability that ${d_ {{\rm{in}}\,\scriptsize +} }$ increases by one is
\begin{equation}
\label{p_d_in+}
\!\!\!P_{{d_ {{\rm{in}}\,\scriptsize +} }}^ +  = \underbrace {\frac{{NL - {d_ {{\rm{in}}\,\scriptsize +} }}}{{NL}}}_{\,\scriptstyle{\rm{select}}\;{\rm{a}}\;{\rm{link}}\;{\rm{which}}\;\hfill\atop
	\scriptstyle{\rm{is}}\;{\rm{not}}\;{\rm{point}}\;{\rm{to}}\;{\rm{Sally}}\hfill}\underbrace {{\pi _S} \cdot \left( {\begin{array}{*{20}{c}}
		{{\alpha _{\scriptsize\overrightarrow { +  + } }}{k_{\scriptsize\overrightarrow { +  + } }}}\\
		{{\alpha _{\scriptsize\overrightarrow { +  - } }}{k_{\scriptsize\overrightarrow { +  - } }}}\\
		{{\alpha _{\scriptsize\overrightarrow { -  + } }}{k_{\scriptsize\overrightarrow { -  + } }}}\\
		{{\alpha _{\scriptsize\overrightarrow { -  - } }}{k_{\scriptsize\overrightarrow { -  - } }}}
		\end{array}} \right)}_{{\rm{break}}\;{\rm{the}}\;{\rm{link}}}\;\underbrace {\frac{1}{{N - 1}}}_{\scriptstyle{\rm{rewire}}\;{\rm{to}}\;{\rm{Sally}}\hfill}.
\end{equation}
On the other hand, Sally is not reconnected provided that her student breaks the selected link. In this case, Sally has one less student. Hence, the transition probability that ${d_ {{\rm{in}}\,\scriptsize +} }$ decreases by one is
\begin{equation}
\label{p_d_in-}
\!P_{{d_ {{\rm{in}}\,\scriptsize +} }}^ -  \!=\!\!\!\!\!\!\!\!\!\!\!\! \underbrace {\frac{{{d_ {{\rm{in}}\,\scriptsize +} }}}{{NL}}}_{\scriptstyle{\rm{select}}\;{\rm{a}}\;{\rm{link}}\;{\rm{which}}\;\hfill\atop
	\scriptstyle{\;\;\rm{is}}\;{\rm{point}}\;{\rm{to}}\;{\rm{Sally}}\hfill}\!\!\!\!\!\!\underbrace {\left( {\frac{{{\pi _{\scriptsize\overrightarrow { +  + } }}{\alpha _{\scriptsize\overrightarrow { +  + } }}{k_{\scriptsize\overrightarrow { +  + } }}}}{{{\pi _{\scriptsize\overrightarrow { +  + } }} + {\pi _{\scriptsize\overrightarrow { -  + } }}}} + \frac{{{\pi _{\scriptsize\overrightarrow { -  + } }}{\alpha _{\scriptsize\overrightarrow { -  + } }}{k_{\scriptsize\overrightarrow { -  + } }}}}{{{\pi _{\scriptsize\overrightarrow { +  + } }} + {\pi _{\scriptsize\overrightarrow { -  + } }}}}} \right)}_{{\rm{break}}\;{\rm{the}}\;{\rm{link}}}\!\underbrace 1_{\scriptstyle{\;\;\rm{rewire}}\;{\rm{to}}\hfill\atop
	\scriptstyle{\rm{other}}\;{\rm{nodes}}\hfill}.
\end{equation}
And $P_{{d_ {{\rm{in}}\,\scriptsize +} }}^0 = 1 - P_{{d_ {{\rm{in}}\,\scriptsize +} }}^ +  - P_{{d_ {{\rm{in}}\,\scriptsize +} }}^ - $ [Fig. \hyperref[tu5]{5}].

\begin{figure}[H]
	\includegraphics[scale=0.25]{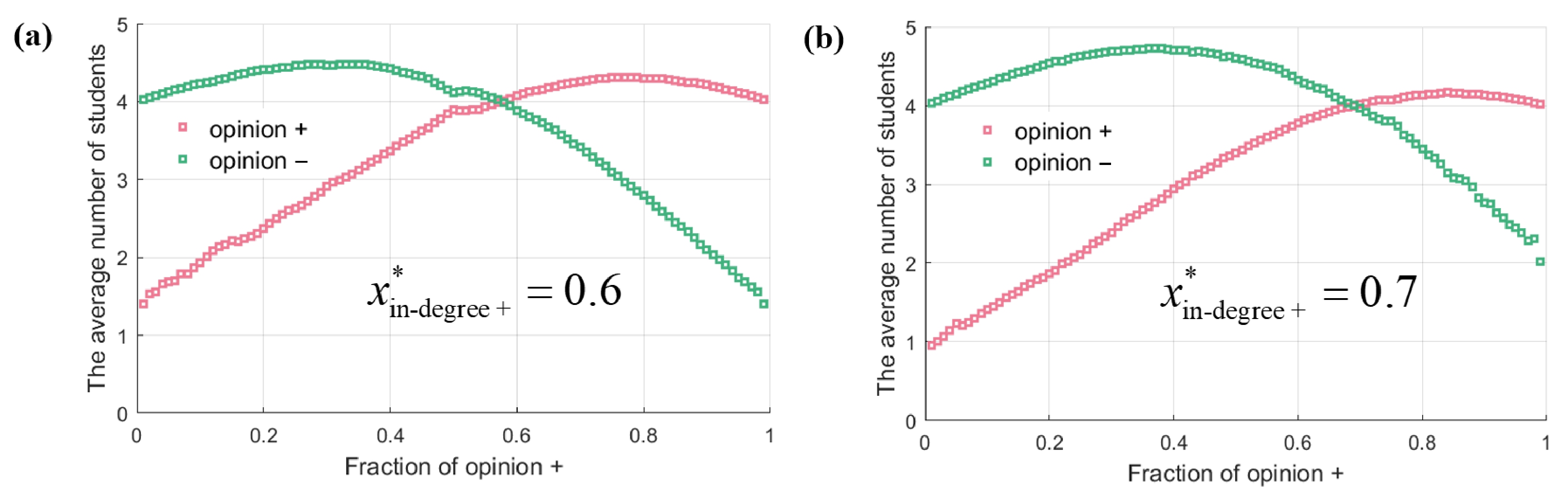}% Here is how to import EPS art
	\caption{\label{tu6}\textbf{Opinion $+$ has as many students as opinion $-$ does in the Nash equilibrium.} For in-group bias, our analysis indicates if the proportion of opinion $+$ is larger than $ x_ {\rm in \mbox{-} degree \kern 1pt \scriptsize +} ^ * $, then the average degree of opinion $+$ is larger than opinion $-$'s. It implies that more students learn opinion $+$. Otherwise, the average degree of opinion $-$ is larger. (a) $x_{\rm{in \mbox{-} degree}{\kern 1pt} \scriptsize + }^ * = 0.6$. Parameters: ${\alpha_{\scriptsize\protect\overrightarrow { X  Y } }} = \alpha = 0.5 $, ${k _{\scriptsize\protect\overrightarrow { +  + } }} = 0.3$, ${k _{\scriptsize\protect\overrightarrow { +  - } }} = 0.9$, ${k _{\scriptsize\protect\overrightarrow { -  + } }} = 0.6$ and ${k _{\scriptsize\protect\overrightarrow { -  - } }} = 0.2$. (b) $x_{\rm{in \mbox{-} degree}{\kern 1pt} \scriptsize + }^ * = 0.7$. Parameters: ${\alpha_{\scriptsize\protect\overrightarrow { X  Y } }} = \alpha = 0.5 $, ${k _{\scriptsize\protect\overrightarrow { +  + } }} = 0.3$, ${k _{\scriptsize\protect\overrightarrow { +  - } }} = 0.6$, ${k _{\scriptsize\protect\overrightarrow { -  + } }} = 0.9$ and ${k _{\scriptsize\protect\overrightarrow { -  - } }} = 0.2$. We run 100 rounds of the simulation. }
\end{figure}

The one-step transition matrix ${P}$ of the Markov process is thus obtained. The Markov chain is aperiodic and irreducible, thus ergodic. Hence it has a unique stationary distribution ${\Xi _D} = \left( {{\xi _0},{\xi _1},{\xi _2}, \cdots {\xi _{N - 1}}} \right)$ which is determined by ${\Xi _D}P = {\Xi _D}$ \cite{KARLIN197545}. Based on \cite{20102}, the stationary distribution is given by
\begin{equation}
\label{eq.12}
\;\;\;\;\;\;{\xi _j} = \frac{{\frac{{P_0^ + }}{{P_j^ - }}\prod\nolimits_{i = 1}^{j - 1} {\frac{{P_i^ + }}{{P_i^ - }}} }}{{1 + \sum\nolimits_{k = 1}^{N - 1} {\frac{{P_0^ + }}{{P_k^ - }}\prod\nolimits_{i = 1}^{k - 1} {\frac{{P_i^ + }}{{P_i^ - }}} } }},\quad 1 \le j \le N - 1
\end{equation}
where the empty product is one, that is, $\prod\nolimits_{i = 1}^0 {{{P_i^ + } \mathord{\left/
			{\vphantom {{P_i^ + } {P_i^ - }}} \right.
			\kern-\nulldelimiterspace} {P_i^ - }}}  = 1$. For $j = 0$, we have ${\xi _0} = $ ${\left( {1 + \sum\nolimits_{k = 1}^{N - 1} {{{P_0^ + } \mathord{\left/
					{\vphantom {{P_0^ + } {P_k^ - }}} \right.
					\kern-\nulldelimiterspace} {P_k^ - }}\prod\nolimits_{i = 1}^{k - 1} {{{P_i^ + } \mathord{\left/
						{\vphantom {{P_i^ + } {P_i^ - }}} \right.
						\kern-\nulldelimiterspace} {P_i^ - }}} } } \right)^{ - 1}}$. When the population size is infinitely large, i.e., $N \to \infty $, we show that the in-degree follows the Poisson distribution (see more details in Supplemental Material). For the average in-degree of opinion $+$, we have $E\left( {{d_ {{\rm{in}}\,\scriptsize +} }} \right) = L{U_+}$, where $L$ is the average in-degree of the network and
\begin{equation}
{U_+} = {{{\pi _S} \cdot \left( {\begin{array}{*{20}{c}}
			{{\alpha _{\scriptsize\overrightarrow { +  + } }}{k_{\scriptsize\overrightarrow { +  + } }}}\\
			{{\alpha _{\scriptsize\overrightarrow { +  - } }}{k_{\scriptsize\overrightarrow { +  - } }}}\\
			{{\alpha _{\scriptsize\overrightarrow { -  + } }}{k_{\scriptsize\overrightarrow { -  + } }}}\\
			{{\alpha _{\scriptsize\overrightarrow { -  - } }}{k_{\scriptsize\overrightarrow { -  - } }}}
			\end{array}} \right)} \mathord{\left/
		{\vphantom {{{\pi _S} \cdot \left( {\begin{array}{*{20}{c}}
						{{\alpha _{\scriptsize\overrightarrow { +  + } }}{k_{\scriptsize\overrightarrow { +  + } }}}\\
						{{\alpha _{\scriptsize\overrightarrow { +  - } }}{k_{\scriptsize\overrightarrow { +  - } }}}\\
						{{\alpha _{\overrightarrow { -  + } }}{k_{\scriptsize\overrightarrow { -  + } }}}\\
						{{\alpha _{\scriptsize\overrightarrow { -  - } }}{k_{\scriptsize\overrightarrow { -  - } }}}
						\end{array}} \right)} {\left( {\displaystyle\frac{{{\pi _{\scriptsize\overrightarrow { +  + } }}{\alpha _{\scriptsize\overrightarrow { +  + } }}{k_{\scriptsize\overrightarrow { +  + } }}}}{{{\pi _{\scriptsize\overrightarrow { +  + } }} + {\pi _{\scriptsize\overrightarrow { -  + } }}}} + \displaystyle\frac{{{\pi _{\scriptsize\overrightarrow { -  + } }}{\alpha _{\scriptsize\overrightarrow { -  + } }}{k_{\scriptsize\overrightarrow { -  + } }}}}{{{\pi _{\scriptsize\overrightarrow { +  + } }} + {\pi _{\scriptsize\overrightarrow { -  + } }}}}} \right)}}} \right.
		\kern-\nulldelimiterspace} {\left( {\displaystyle\frac{{{\pi _{\scriptsize\overrightarrow { +  + } }}{\alpha _{\scriptsize\overrightarrow { +  + } }}{k_{\scriptsize\overrightarrow { +  + } }}}}{{{\pi _{\scriptsize\overrightarrow { +  + } }} + {\pi _{\scriptsize\overrightarrow { -  + } }}}} + \displaystyle\frac{{{\pi _{\scriptsize\overrightarrow { -  + } }}{\alpha _{\scriptsize\overrightarrow { -  + } }}{k_{\scriptsize\overrightarrow { -  + } }}}}{{{\pi _{\scriptsize\overrightarrow { +  + } }} + {\pi _{\scriptsize\overrightarrow { -  + } }}}}} \right)}}.
\end{equation}
Interestingly, ${U_ {\scriptsize +} } = {{{g_ {\scriptsize +} }} \mathord{\left/
		{\vphantom {{{g_ {\scriptsize +} }} {\left( {{x_ {\scriptsize +} }{g_ {\scriptsize +} } + {x_ {\scriptsize -} }{g_ {\scriptsize -} }} \right)}}} \right.
		\kern-\nulldelimiterspace} {\left( {{x_ {\scriptsize +} }{g_ {\scriptsize +} } + {x_ {\scriptsize -} }{g_ {\scriptsize -} }} \right)}}$, where ${g_ {\scriptsize +} }$(${g_ {\scriptsize -} }$) is regarded as the payoff of the opinion $+$($-$). Hence, the expectation of in-degree for the two opinions is fully captured by an emergent three-player two-strategy game (See Supplemental Material for details), whose payoff table is given by
\begin{table}[H]
	\centering
	\caption{\label{table 4}The transient topology is captured by the emergent payoff matrix of three-player two-strategy game. }
%	\begin{ruledtabular}
		\begin{tabular}{cccc}
			\br
			Individual(s)&\qquad2+&\qquad1+&\qquad0+\\
			%\mbox{Three}&\mbox{Four}&\mbox{Five}\\
			\br
			+&\qquad$a_1$&\qquad\mbox{$a_2/2$}&\qquad\mbox{$a_3$}\\
			$-$&\qquad$b_1$&\qquad$b_2/2$&\qquad$b_3$\\
			\br
		\end{tabular}
%	\end{ruledtabular}
\end{table}
\noindent where ${a_1} = {{{\alpha _{\scriptsize\overrightarrow { +  - } }}{\beta _{\scriptsize\overrightarrow { -  + } }}} \mathord{\left/
		{\vphantom {{{\alpha _{\scriptsize\overrightarrow { +  - } }}{\beta _{\scriptsize\overrightarrow { -  + } }}} {{k_{\scriptsize\overrightarrow { +  + } }}}}} \right.
		\kern-\nulldelimiterspace} {{k_{\scriptsize\overrightarrow { +  + } }}}}, {b_1} = {{{\alpha _{\scriptsize\overrightarrow { +  + } }}{\beta _{\scriptsize\overrightarrow { -  + } }}} \mathord{\left/
		{\vphantom {{{\alpha _{\scriptsize\overrightarrow { +  + } }}{\beta _{\scriptsize\overrightarrow { -  + } }}} {{k_{\scriptsize\overrightarrow { +  - } }}}}} \right.
		\kern-\nulldelimiterspace} {{k_{\scriptsize\overrightarrow { +  - } }}}}, {a_2} = {{{\alpha _{\scriptsize\overrightarrow { +  - } }}{\beta _{\scriptsize\overrightarrow { +  + } }}} \mathord{\left/
		{\vphantom {{{\alpha _{\scriptsize\overrightarrow { +  - } }}{\beta _{\scriptsize\overrightarrow { +  + } }}} {{k_{\scriptsize\overrightarrow { -  + } }}}}} \right.
		\kern-\nulldelimiterspace} {{k_{\scriptsize\overrightarrow { -  + } }}}} + {{\left[ {{\alpha _{\scriptsize\overrightarrow { -  + } }}\left( {{\alpha _{\scriptsize\overrightarrow { +  - } }} - {\alpha _{\scriptsize\overrightarrow { -  - } }}} \right) + } \right.}} $\\$\alpha _{\scriptsize\overrightarrow { -  - } }\beta _{\scriptsize\overrightarrow { +  - } }]/k_{\scriptsize\overrightarrow { +  + } },  {b_2} = {{{\alpha _{\scriptsize\overrightarrow { -  + } }}{\beta _{\scriptsize\overrightarrow { -  - } }}} \mathord{\left/
		{\vphantom {{{\alpha _{\scriptsize\overrightarrow { -  + } }}{\beta _{\scriptsize\overrightarrow { -  - } }}} {{k_{\scriptsize\overrightarrow { +  - } }}}}} \right.
		\kern-\nulldelimiterspace} {{k_{\scriptsize\overrightarrow { +  - } }}}} + {{\left[ {{\alpha _{\scriptsize\overrightarrow { -  + } }}\left( {{\alpha _{\scriptsize\overrightarrow { +  - } }} - {\alpha _{\scriptsize\overrightarrow { +  + } }}} \right) + {\alpha _{\scriptsize\overrightarrow { +  + } }}{\beta _{\scriptsize\overrightarrow { +  - } }}} \right]} \mathord{\left/
		{\vphantom {{\left[ {{\alpha _{\scriptsize\overrightarrow { -  + } }}\left( {{\alpha _{\scriptsize\overrightarrow { +  - } }} - {\alpha _{\scriptsize\overrightarrow { +  + } }}} \right) + {\alpha _{\scriptsize\overrightarrow { +  + } }}{\beta _{\scriptsize\overrightarrow { +  - } }}} \right]} {{k_{\scriptsize\overrightarrow { -  - } }}}}} \right.
		\kern-\nulldelimiterspace} {{k_{\scriptsize\overrightarrow { -  - } }}}}$, ${a_3} = {{{\alpha _{\scriptsize\overrightarrow { -  - } }}{\beta _{\scriptsize\overrightarrow { +  - } }}} \mathord{\left/
		{\vphantom {{{\alpha _{\scriptsize\overrightarrow { -  - } }}{\beta _{\scriptsize\overrightarrow { +  - } }}} {{k_{\scriptsize\overrightarrow { -  + } }}}}} \right.
		\kern-\nulldelimiterspace} {{k_{\scriptsize\overrightarrow { -  + } }}}}, {b_3} = {{{\alpha _{\scriptsize\overrightarrow { -  + } }}{\beta _{\scriptsize\overrightarrow { +  - } }}} \mathord{\left/
		{\vphantom {{{\alpha _{\scriptsize\overrightarrow { -  + } }}{\beta _{\scriptsize\overrightarrow { +  - } }}} {{k_{\scriptsize\overrightarrow { -  - } }}}}} \right.
		\kern-\nulldelimiterspace} {{k_{\scriptsize\overrightarrow { -  - } }}}}$. The Nash equilibrium of the emergent game is the transient topology, at which the two opinions have the same student size [Fig. \hyperref[tu6]{6}]. If the payoff of opinion $+$ is larger than the payoff of opinion $-$ for \hyperref[table 4]{Table 4}, then the average in-degree of opinion $+$ is greater than that of opinion $-$.

\begin{figure}
	\centering
	\includegraphics[scale=0.35]{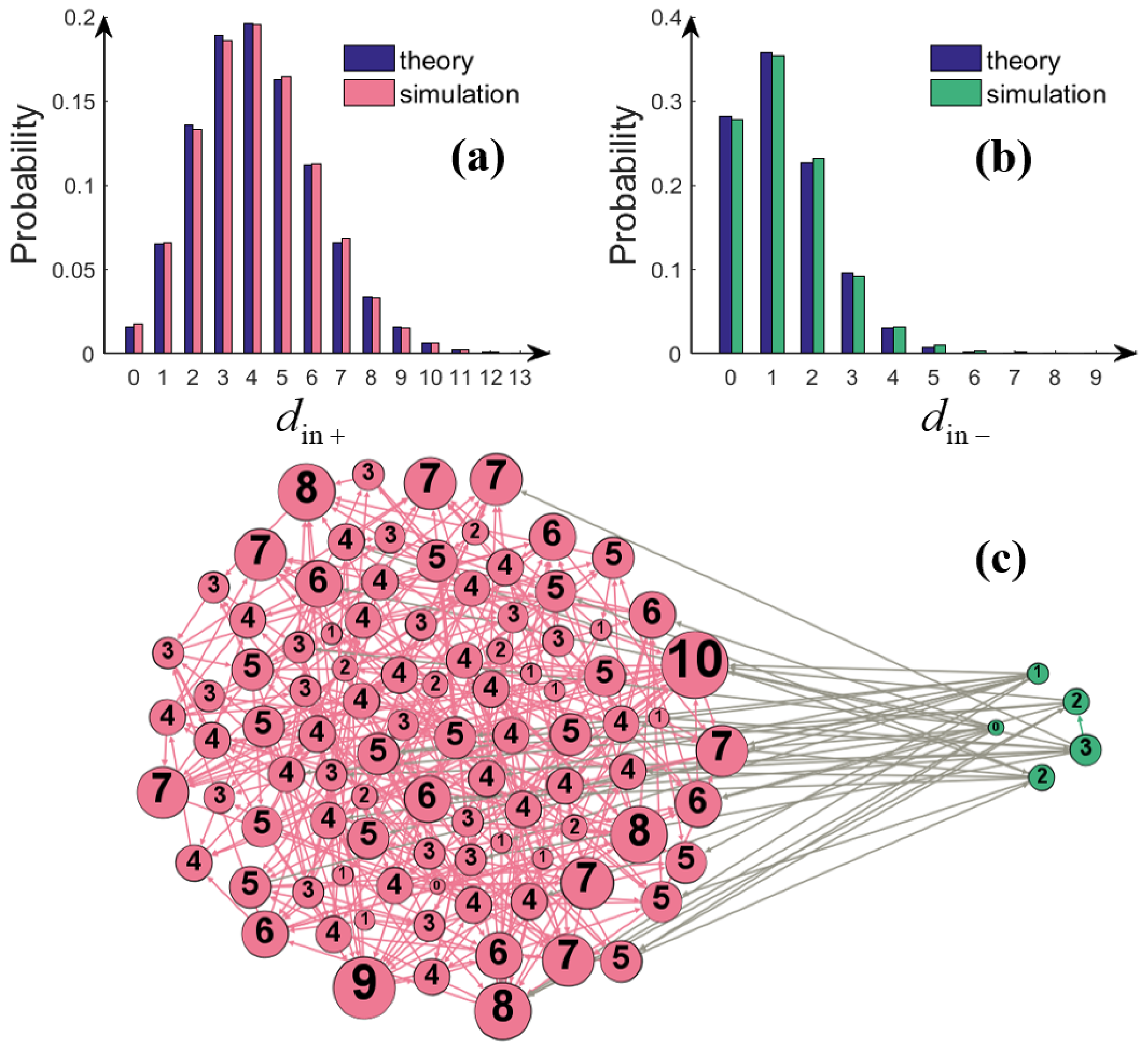}% Here is how to import EPS art
	\caption{\label{tu7} \textbf{Student size distribution.} The accuracy of our model with respect to the simulation results is good shown in (a) and (b), even though our analysis ignores the out-degree. For opinion $+$, the theoretical value of the average in-degree is 4.1428 and the simulation value is 4.1422. The order of magnitude of the error is ${10^{ - 4}}$. For opinion $-$, the theoretical value of the average in-degree is 1.2672 and the simulation value is 1.2985. The order of magnitude of the error is ${10^{ - 2}}$. (c) Transient topology of the directed network. The red nodes take opinion $+$ and the green nodes take opinion $-$. The label and the size of the nodes are determined by their in-degree. When ${x_ {\scriptsize +} }$ is large, the nodes with opinion $-$ have a low in-degree, which implies that the number of students whose teachers hold the opinion $-$ is small. As a result, it is difficult to spread the opinion $-$, and opinion $-$ ultimately invades unsuccessfully. When ${x_ {\scriptsize -} }$ is small, the connections between opinion $-$ are not strong. As shown in the figure, there is only one directed green edge between opinion $-$. The visualization of the directed network is obtained via GEPHI \cite{GEPHI}. Parameters: ${k_{\scriptsize\protect\overrightarrow { X  Y } }} = k = 0.5 $, ${\alpha _{\scriptsize\protect\overrightarrow { +  + } }} = 0.1$, ${\alpha _{\scriptsize\protect\overrightarrow { +  - } }} = 0.6$, ${\alpha _{\scriptsize\protect\overrightarrow { -  + } }} = 0.6$ and ${\alpha _{\scriptsize\protect\overrightarrow { -  - } }} = 0.1$. The initial state is $x_ {\scriptsize +} = 0.5$. When ${x_ {\scriptsize +} } = 0.95$, we record the number of students who adopt either opinion $+$ or opinion $-$ for each individual. We run ${10^3}$ rounds of the simulation and set $N = 100, L = 4$, and $w = 0.01$.}
\end{figure}

When the four probabilities of breaking links are the same, i.e., ${k_{\scriptsize\overrightarrow {X Y} }} = k $, where $\overrightarrow {XY}  \in S$ and $0<k<1$ and concentrate on the in-group bias. Noteworthily, Eq. \hyperref[p_d_in+]{(10)} and Eq. \hyperref[p_d_in-]{(11)} are approximations because Sally's out-degree i.e., her teachers are neglected. Bidirectional links are not excluded in the approximation. In spite of this error, the in-degree distribution via the simulation agrees perfectly with the theoretical approximations for both one opinion in majority and the other opinion in minority [Figs. \hyperref[tu7]{7(a)} and \hyperref[tu7]{7(b)}]. Intuitively, here $N \gg L$, i.e., the total number of individuals is much larger than the number of students for an individual which is close to the reality. Thus, each node almost obeys the same in-degree distribution and each update is approximately independent. Hence, these approximations are acceptable. For the completeness of our study, we show the corresponding results for the out-degree (See Supplemental Material for details).

\subsection{Emergent two-player games: the student size}
We focus on two classes of breaking patterns, i.e., the probability of choosing nodes ${\alpha_{\scriptsize\overrightarrow {XY} }}$ and the probability of breaking directed links ${k_{\scriptsize\overrightarrow {XY} }}$. If there exists $0<\alpha<1$, s.t., ${\alpha _{\scriptsize\overrightarrow {XY} }} = \alpha $, the three-player game degenerates to the two-player game
\begin{equation}
\label{eq.14}
{M_{{\rm{in \mbox{-} degree}}}} = \left( {\begin{array}{*{20}{c}}
	{\displaystyle\frac{1}{{{k_{\scriptsize\overrightarrow { +  + } }}}}}&{\displaystyle\frac{1}{{{k_{\scriptsize\overrightarrow { -  + } }}}}}\\
	{\displaystyle\frac{1}{{{k_{\scriptsize\overrightarrow { +  - } }}}}}&{\displaystyle\frac{1}{{{k_{\scriptsize\overrightarrow { -  - } }}}}}
	\end{array}} \right).
\end{equation}

We obtain a internal equilibrium $x_{\rm{in \mbox{-} degree} \kern 1pt \scriptsize + }^ * $ for in-group bias
\begin{equation}
x_{\rm{in \mbox{-} degree} \kern 1pt \scriptsize + }^ * = \displaystyle\frac{{{1 \mathord{\left/
				{\vphantom {1 {{k_{\scriptsize\overrightarrow { -  - } }}}}} \right.
				\kern-\nulldelimiterspace} {{k_{\scriptsize\overrightarrow { -  - } }}}} - {1 \mathord{\left/
				{\vphantom {1 {{k_{\scriptsize\overrightarrow { -  + } }}}}} \right.
				\kern-\nulldelimiterspace} {{k_{\scriptsize\overrightarrow { -  + } }}}}}}{{{1 \mathord{\left/
				{\vphantom {1 {{k_{\scriptsize\overrightarrow { +  + } }}}}} \right.
				\kern-\nulldelimiterspace} {{k_{\scriptsize\overrightarrow { +  + } }}}} - {1 \mathord{\left/
				{\vphantom {1 {{k_{\scriptsize\overrightarrow { +  - } }}}}} \right.
				\kern-\nulldelimiterspace} {{k_{\scriptsize\overrightarrow { +  - } }}}} - {1 \mathord{\left/
				{\vphantom {1 {{k_{\scriptsize\overrightarrow { -  + } }}}}} \right.
				\kern-\nulldelimiterspace} {{k_{\scriptsize\overrightarrow { -  + } }}}} + {1 \mathord{\left/
				{\vphantom {1 {{k_{\scriptsize\overrightarrow { -  - } }}}}} \right.
				\kern-\nulldelimiterspace} {{k_{\scriptsize\overrightarrow { -  - } }}}}}}.
\end{equation}

The equilibrium is a Nash equilibrium of the emergent game Eq. \hyperref[eq.14]{(14)}. It refers to a \emph{topology} in which opinion $+$ has as many students as opinion $-$ does [Fig. \hyperref[tu6]{6}]. For in-group bias, if ${x_ {\scriptsize +} } > x_ {\rm in \mbox{-} degree \kern 1pt \scriptsize +} ^ * $, the average degree of opinion $+$ is larger than opinion $-$'s. It implies that more students learn opinion $+$. Otherwise, the average degree of opinion $-$ is larger.

\begin{figure}
	\includegraphics[scale=0.28]{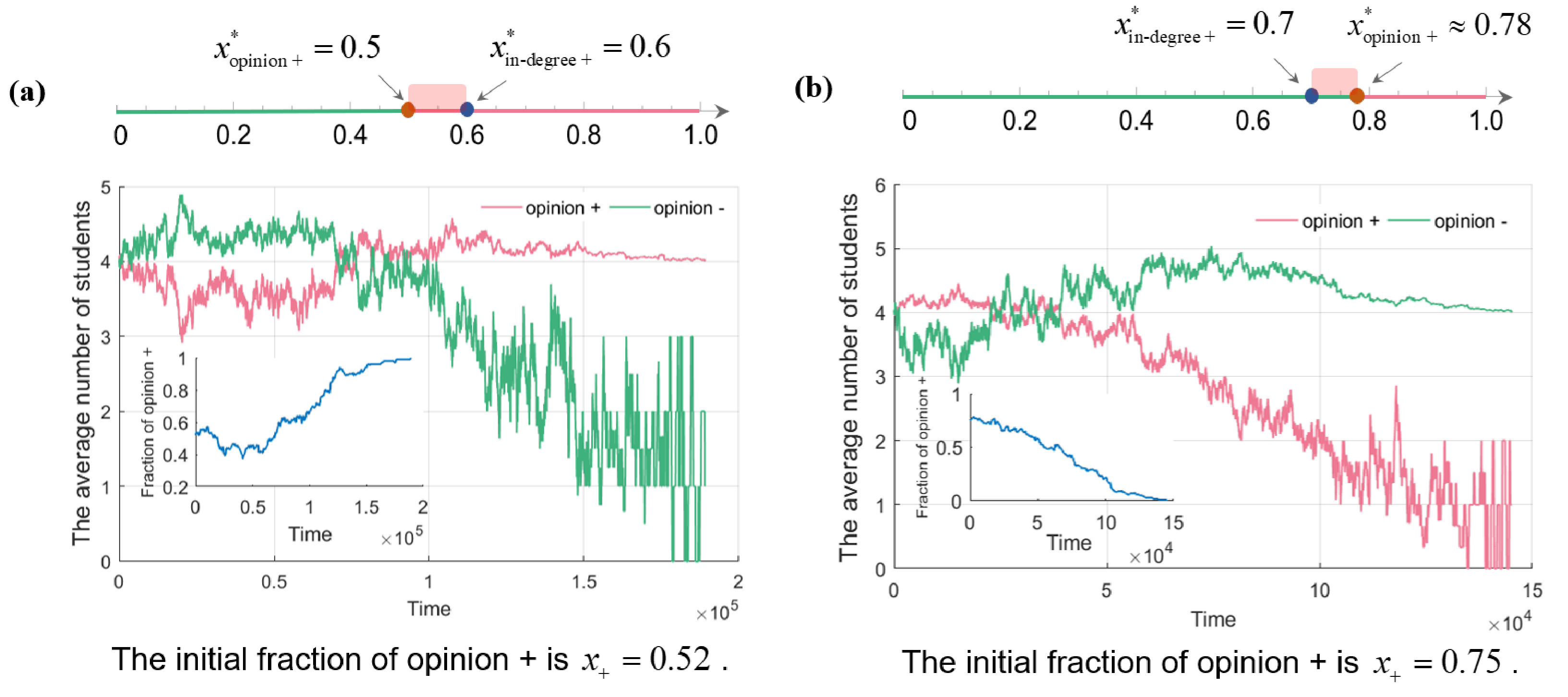}% Here is how to import EPS art
	\caption{\label{tu8}\textbf{Opinions with fewer students can invade successfully.} For in-group bias, the number of disciples of an opinion is not the key factor in the success of invasion. The opinion with a large number of students does not necessarily end up with a successful invasion, and that with a small number of students does not necessarily invade unsuccessfully. (a) Even if more students learn the opinion $-$ than the opinion $+$, the opinion $+$ eventually wins. Parameters: ${\alpha_{\scriptsize\protect\overrightarrow { X  Y } }} = \alpha = 0.5 $, ${k _{\scriptsize\protect\overrightarrow { +  + } }} = 0.3$, ${k _{\scriptsize\protect\overrightarrow { +  - } }} = 0.9$, ${k _{\scriptsize\protect\overrightarrow { -  + } }} = 0.6$ and ${k _{\scriptsize\protect\overrightarrow { -  - } }} = 0.2$. $x_{\rm{opinion}{\kern 1pt} \scriptsize + }^ * = 0.5$ and $x_{\rm{in \mbox{-} degree}{\kern 1pt} \scriptsize + }^ * = 0.6$. The initial fraction of opinion + is 0.52. (b) Even if more students learn the opinion $+$ than the opinion $-$, the opinion $-$ eventually wins. Parameters: ${\alpha_{\scriptsize\protect\overrightarrow { X  Y } }} = \alpha = 0.5 $, ${k _{\scriptsize\protect\overrightarrow { +  + } }} = 0.3$, ${k _{\scriptsize\protect\overrightarrow { +  - } }} = 0.6$, ${k _{\scriptsize\protect\overrightarrow { -  + } }} = 0.9$ and ${k _{\scriptsize\protect\overrightarrow { -  - } }} = 0.2$. $x_{\rm{in \mbox{-} degree}{\kern 1pt} \scriptsize + }^ * = 0.7$ and $x_{\rm{opinion}{\kern 1pt} \scriptsize + }^ * \approx 0.78$. The initial fraction of opinion + is 0.75. }
\end{figure}

Since the emergent games $M_{\rm{opinion}}$, i.e., Eq. \hyperref[eq.4]{(4)} and $M_{\rm{in \mbox{-} degree}}$, i.e., Eq. \hyperref[eq.14]{(14)} are not equal, we cannot capture both the opinion formation and the transient topology with just one emergent game. Thus, here are some counterintuitive cases. For in-group bias, ${k _{\scriptsize\overrightarrow { +  - } }} > {k _{\scriptsize\overrightarrow { +  + } }}$ and $ {k _{\scriptsize\overrightarrow { -  + } }} > {k _{\scriptsize\overrightarrow { -  - } }}$, if the initial proportion of opinion $+$ is larger than $x_ {\rm opinion \kern 1pt \scriptsize +} ^ *$, then opinion $+$ is likely to take over. For ${k_{\scriptsize\overrightarrow { +  - } }} > {k_{\scriptsize\overrightarrow { -  + } }}$, we have $x_{\rm{opinion}{\kern 1pt} \scriptsize + }^ * < x_{\rm{in \mbox{-} degree}{\kern 1pt} \scriptsize + }^ *$. If the initial fraction of opinion $+$ is between $x_ {\rm opinion \kern 1pt \scriptsize +} ^ *$ and $x_ {\rm in \mbox{-} degree \kern 1pt \scriptsize +} ^ *$, then opinion $+$ invades successfully in the end, even if more students learn the opinion $-$ than opinion $+$ in the beginning [Fig. \hyperref[tu8]{8(a)}]. Similarly, if ${k_{\scriptsize\overrightarrow { +  - } }} < {k_{\scriptsize\overrightarrow { -  + } }}$, we have $x_{\rm{in \mbox{-} degree}{\kern 1pt} \scriptsize + }^ * < x_{\rm{opinion}{\kern 1pt} \scriptsize + }^ *$. And if the initial fraction of opinion $+$ is between $x_ {\rm in \mbox{-} degree \kern 1pt \scriptsize +} ^ *$ and $x_ {\rm opinion \kern 1pt \scriptsize +} ^ *$, then opinion $+$ invades unsuccessfully eventually, even if more students learn the opinion $+$ than the opinion $-$ in the beginning [Fig. \hyperref[tu8]{8(b)}]. It implies that the opinion with few students is likely to invade successfully. Hence, the student size is not the indicator of the successful invasion, which is counterintuitive.

If there is $0<k<1$, s.t., ${k _{\scriptsize\overrightarrow {XY} }} = k $, then the degenerated payoff matrix is
\begin{equation}
\label{eq.15}
{R_{{\rm{in \mbox{-} degree}}}} = \left( {\begin{array}{*{20}{c}}
	{{\alpha _{\scriptsize\overrightarrow { +  - } }}{\beta _{\scriptsize\overrightarrow { -  + } }}}&{{\alpha _{\scriptsize\overrightarrow { -  - } }}{\beta _{\scriptsize\overrightarrow { +  - } }}}\\
	{{\alpha _{\scriptsize\overrightarrow { +  + } }}{\beta _{\scriptsize\overrightarrow { -  + } }}}&{{\alpha _{\scriptsize\overrightarrow { -  + } }}{\beta _{\scriptsize\overrightarrow { +  - } }}}
	\end{array}} \right).
\end{equation}
$R_ {\rm{in \mbox{-} degree}}$ is the same as $R_ {\rm opinion}$, i.e., Eq. \hyperref[eq.6]{(6)}. It implies that the internal equilibrium $y_{\rm{in \mbox{-} degree} \kern 1pt \scriptsize + }^ * $ is equal to $y_{\rm opinion \kern 1pt  \scriptsize +}^ *$. For the in-group bias, if the initial fraction of opinion $+$ is larger than $y_{\rm opinion \kern 1pt  \scriptsize +}^ *$, then more students learn the opinion $+$ than the opinion $-$ and the opinion $+$ invades successfully. It implies that the student size is the indicator of the successful invasion in this case. We draw the directed network topology [Fig. \hyperref[tu7]{7(c)}]. If the proportion of one opinion is quite small, then the in-degree of the opinion is small.

\subsection{An emergent three-player game for the student size: complexity analysis}
Some of the three-player games may be expanded by the two-player games. We take the number of internal equilibria of the replicated equation as the true complexity of our model. Based on Descartes’ rule of signs \cite{2008Polynomials}, if ${a_1}>{b_1}$, ${a_2}<{b_2}$ and ${a_3}>{b_3}$ (or ${a_1}<{b_1}$, ${a_2}>{b_2}$ and ${a_3}<{b_3}$ ) are satisfied in \hyperref[table 4]{Table 4}, the three-player two-strategy game has at most two internal equilibria. At the equilibria, the in-degree of opinion $+$ and that of opinion $-$ are equal. To verify whether the same parameters simultaneously lead to three internal equilibria in a four-player two-strategy game and two internal equilibria in a three-player two-strategy game, we take the set of parameters, i.e., ${k_{\scriptsize\overrightarrow { +  + } }} = \rho ,{k_{\scriptsize\overrightarrow { +  - } }} = \rho ,{k_{\scriptsize\overrightarrow { -  + } }} = \rho/4,{k_{\scriptsize\overrightarrow { -  - } }} = \rho ,{\alpha _{\scriptsize\overrightarrow { +  + } }} = \rho/2,{\alpha _{\scriptsize\overrightarrow { +  - } }} = \rho ,{\alpha _{\scriptsize\overrightarrow { -  + } }} = 2\rho$ and ${\alpha _{\scriptsize\overrightarrow { -  - } }} = \rho/4$, where $0 < \rho  < 0.5$ into \hyperref[table 4]{Table 4}. However, there is only one internal equilibrium. This emergent three-player game differs in complexity from the four-player game to predict the fate of opinions. We show that the four-player two-strategy game with three internal equilibria and the three-player two-strategy game with two internal equilibria cannot occur at the same time (Supplemental Material). It indicates that the complexity of the two emergent games is different and we can not use the same emergent game to describe both the fate of opinions and the transient topology except some special cases, i.e., ${k _{\scriptsize\overrightarrow {XY} }} = k $, where $0<k<1$. We find a new set of parameters in which the three-player two-strategy game has two internal equilibria, as shown in Supplemental Material.

\section{Conclusion and Discussion}
Evolutionary game theory is a powerful mathematical framework to explore how individuals adjust their strategies, provided that the game interactions are given in prior \cite{b11, b12, b13}. Both opinion dynamics and evolutionary game dynamics have been benefited from the statistical physics method, yet they are treated as two distinct fields. We show that opinion dynamics is equivalent to the evolutionary games, both opinion wise and network wise. We focus on a voter model on an evolving directed network without any game interactions. We have shown that the fate of opinions is captured by a replicator equation of an emergent four-player two-strategy game. The complexity of the fate of opinions is thus the same as the classic evolutionary four-player two-strategy game. It has at most three internal equilibria. This equivalence result explicitly captures how opinions reach a consensus and how opinions coexist for a long time, which are the two main questions in opinion dynamics. On the other hand, we show that the transient topology is fully captured by an emergent three-player two-strategy game. Thus it has at most two internal equilibria. The Nash equilibrium of the emergent game is the transient topology, at which the two opinions have the same student size. We obtain the in(out)-degree distribution, which is typically challenging in previous works. This equivalence result explicitly tells who has how many neighbors during the opinion formation. Thus it demonstrates the transient topology during opinion formation.

The emergent games degenerate to two-player two-strategy games, if the type of directed links is not considered when selecting an individual or initiating breaking the link, i.e., ${\alpha _{\scriptsize\overrightarrow {XY} }} = \alpha $ or ${k _{\scriptsize\overrightarrow {XY} }} = k $, where $0<\alpha<1$, $0<k<1$ and $\overrightarrow {XY}  \in S$. If we focus on the bi-directionality and set ${\alpha _{\scriptsize\overrightarrow {XY} }} = \alpha = 1/2 $, the emergent game which captures the fate of opinions, i.e., Eq. \hyperref[eq.4]{(4)} is equivalent to \cite{Bridging} where networks are undirected yet dynamical. For in-group bias, individuals can reach a consensus. For out-group bias, opinions can coexist if opinions coexist in the beginning. Furthermore, the condition ${\alpha _{\scriptsize\overrightarrow {XY} }} = \alpha $ can be relaxed to ${\alpha _{\scriptsize\overrightarrow { +  + } }} = {\alpha _{\scriptsize\overrightarrow { -  + } }} = {\gamma _1}$ and ${\alpha _{\scriptsize\overrightarrow { +  - } }} = {\alpha _{\scriptsize\overrightarrow { -  - } }} = {\gamma _2}$, where $0<{\gamma _1}, {\gamma _2}<1$. For example, if the teachers have the same opinion $+$, then their students have the same probability of being selected, i.e., ${\alpha _{\scriptsize\overrightarrow { +  + } }} = {\alpha _{\scriptsize\overrightarrow { -  + } }} = {\gamma _1}$. We have
\begin{equation}
\begin{array}{l}
M_{\rm {opinion\_new} } = \begin{array}{*{20}{c}}
{\left( {\begin{array}{*{20}{c}}
		{\displaystyle\frac{{\gamma _2} {k_{\scriptsize\overrightarrow { +  - } }}}{{{{\gamma _1} k_{\scriptsize\overrightarrow { +  + } }}}}}&1\\
		1&{\displaystyle\frac{{{\gamma _1} k_{\scriptsize\overrightarrow { -  + } }}}{{{{\gamma _2} k_{\scriptsize\overrightarrow { -  - } }}}}}
		\end{array}} \right)}
\end{array}
\end{array}
\end{equation}
and
\begin{equation}
{M_{{\rm{in \mbox{-} degree\_new}}}} = \left( {\begin{array}{*{20}{c}}
	{\displaystyle\frac{{\gamma _2}}{{{k_{\scriptsize\overrightarrow { +  + } }}}}}&{\displaystyle\frac{{\gamma _2}}{{{k_{\scriptsize\overrightarrow { -  + } }}}}}\\
	{\displaystyle\frac{{\gamma _1}}{{{k_{\scriptsize\overrightarrow { +  - } }}}}}&{\displaystyle\frac{{\gamma _1}}{{{k_{\scriptsize\overrightarrow { -  - } }}}}}
	\end{array}} \right).
\end{equation}
If ${\gamma _1} = {\gamma _2}$, then $M_{\rm opinion } = M_{\rm opinion\_new } $ and ${M_{{\rm{in \mbox{-} degree}}}} = {M_{{\rm{in \mbox{-} degree\_new}}}}$.

We reveal a counterintuitive phenomenon with the aid of the two different emergent games, i.e., $M_{\rm{opinion}}$ [Eq. \hyperref[eq.4]{(4)}] and $M_{\rm{in \mbox{-} degree}}$ [Eq. \hyperref[eq.14]{(14)}]. Intuitively, if the number of disciples of opinion $+$ is larger than the opinion $-$, then opinion $+$ is learned by more students, hence the fraction of opinion $+$ increases and opinion $+$ can take over the whole population. However, we show that the number of disciples is \emph{not} the key to the success of the invasion. An opinion with a smaller student size can succeed in the population. Noteworthily, if ${k_{\scriptsize\overrightarrow { +  - } }} = {k_{\scriptsize\overrightarrow { -  + } }} = k$, where $0<k<1$, we have $M_{\rm{opinion}} = k \cdot M_{\rm{in \mbox{-} degree}}$. It implies that one emergent game is sufficient to capture both the fate of opinions and the transient topology. We also show ${M_{{\rm{in \mbox{-} degree}}}}$ is the same as ${M_{{\rm{out \mbox{-} degree}}}}$ in this case (See Supplemental Material). It implies that the average in-degree is equal to the average out-degree, i.e., one individual has the same number of students and teachers on average. It mirrors an undirected-like network. In other words, if we do not distinguish $\overrightarrow { +  - } $ and $\overrightarrow { -  + } $, the network has symmetric-like properties in a statistical sense although it is still a directed network. Furthermore, the number of students with popular opinions is not higher than that with non-popular opinions, whereas opinion leaders play a decisive role in static networks \cite{PhysRevE.81.057103}. It implies that undirected and directed networks are fundamentally different.

Clustering is believed to play a crucial role in complex systems \cite{PMID:28039984, recipe, CZAPLICKA2022112363, doi:10.1073/pnas.2113468118, Ohtsuki2006A, Asymmetric}. However, we find that if individuals with opinion $+$ gather together, the opinion $+$ does not necessarily invade successfully. It implies that the clustering of individuals with the same opinions is \emph{not} the key to a successful invasion in the dynamical directed network (see more details in Supplemental Material).

To sum up, our work bridges the gap between the opinion dynamics and evolutionary game theory. Via the bridge, we are able to predict both the fate of opinions and the transient topology from a game perspective.

\section*{Acknowledgments}
We gratefully acknowledge Xunlong Wang, who inspire us to find that the in-degree follows the Poisson distribution in the infinite large population size limit. We appreciate NSFC No.61751301.

\appendix

\section{Linking Dynamics} \label{Appendix A}
Here the number of directed links $NL$ is constant. Each directed link
$i \left( {i = 1,2, \cdots ,NL} \right)$ is selected with probability $1/NL$. In time $t$, we randomly select a directed link ${i^t} = i$. If the selected ${i^t}$ does not break, then we have ${i^{t + 1}} = {i^t}$. Otherwise, a new directed link is introduced, denoted as ${i^{t + 1}}$. We denote the type of directed edge of ${i^t}$ by $T\left( {{i^t}} \right)$, where
$T\left( {{i^t}} \right) \in S$.

The linking dynamics is captured by Markov chain with transition matrix ${Q_{\scriptsize ( {\overrightarrow {AB} } ) ( {\overrightarrow {CD} } )}}$, which is the probability that link $\overrightarrow {AB} $ transforms to link $\overrightarrow {CD} $ in one time step. For instance, ${Q_{\scriptsize ( {\overrightarrow { +  - } } ) ( {\overrightarrow { +  + } } ) }}$ is the probability that ${i^t}$ of type $\overrightarrow { +  - } $ transforms to ${i^{t + 1}}$ of type $\overrightarrow { +  + } $. In this case, one of the following two cases occurs:\\
(1) ${i^t}$ is not selected (with probability $\left( {NL - 1} \right)/NL$).\\
(2) \!${i^t}$ \!is selected (with probability $1/NL$). Then, either the original $\overrightarrow { +  - } $ link is not broken (with probability $1 - {k_{\scriptsize\overrightarrow { +  - } }}$) or the selected student with opinion $+$ reconnects a new teacher with opinion $+$ when the original $\overrightarrow { +  - } $ link is broken (with probability ${{k_{\scriptsize\overrightarrow { +  - } }}\alpha _{\scriptsize\overrightarrow { +  - } }} x_{\scriptsize +}$, where $x_{\scriptsize +}$ is the fraction of opinion $+$). Hence,

\begin{equation}
\setcounter{equation}{1}
\renewcommand\theequation{A\arabic{equation}}
{Q_{\left( {\scriptsize\overrightarrow { +  - } } \right)\left( {\scriptsize\overrightarrow { +  + } } \right)}} = \frac{{NL - 1}}{{NL}} + \frac{1}{{NL}}\left( {1 - {k_{\scriptsize\overrightarrow { +  - } }} + {k_{\scriptsize\overrightarrow { +  - } }}{\alpha _{\scriptsize\overrightarrow { +  - } }}{x_ {\scriptsize +} }} \right).
\end{equation}
And $x_{\scriptsize -}=1-x_{\scriptsize +}$ is the fraction of opinion $-$. The transition probability matrix is given by
\begin{equation}
\setcounter{equation}{2}
\renewcommand\theequation{A\arabic{equation}}
Q = \frac{{NL - 1}}{{NL}}{I_4} + \frac{1}{{NL}}V,
\end{equation}
where ${I_4}$ is the identity matrix and the matrix $V$ is given by Eq. \hyperref[pingwenfenbu]{(A.3)}.\\
$V = $
%\begin{widetext}
	%\begin{small}
	\setlength{\arraycolsep}{0.6pt}
	\begin{equation}
	\hspace{-38mm}%{}
	\begin{array}{*{20}{c}}
	\quad{\begin{array}{*{20}{c}}
		{}&{\begin{array}{*{20}{c}}
			{\begin{array}{*{20}{c}}
				{\begin{array}{*{20}{c}}
					\quad\quad\quad\quad{\overrightarrow { +  + } }
					\end{array}}\quad\quad\quad\quad\quad\quad\quad\quad\;
				\end{array}} {\overrightarrow { +  - } }
			\end{array}\;\quad\quad\quad\quad\quad\quad\quad\quad\quad
			\begin{array}{*{20}{c}}\,
			{\overrightarrow { -  + } }\;\quad\quad\quad\quad\quad\quad\quad\quad\;\;\,{\overrightarrow { -  - } }\quad\quad\quad\quad\quad
			\end{array}}\\
		{\begin{array}{*{20}{c}}
			{\overrightarrow { +  + } }\\
			{\overrightarrow { +  - } }\\
			{\overrightarrow { -  + } }\\
			{\overrightarrow { -  - } }
			\end{array}}&{\left( {\begin{array}{*{20}{c}}
				{1 - {k_{\scriptsize\overrightarrow { +  + } }} + {k_{\scriptsize\overrightarrow { +  + } }}{x_{\scriptsize +} }}&{{k_{\scriptsize\overrightarrow { +  + } }}{\alpha _{\scriptsize\overrightarrow { +  + } }}{x_{\scriptsize -} }}&{{k_{\scriptsize\overrightarrow { +  + } }}{\beta _{\scriptsize\overrightarrow { +  + } }}{x_{\scriptsize -} }}&0\\
				{{k_{\scriptsize\overrightarrow { +  - } }}{\alpha _{\scriptsize\overrightarrow { +  - } }}{x_{\scriptsize +} }}&{1 - {k_{\scriptsize\overrightarrow { +  - } }}{\alpha _{\scriptsize\overrightarrow { +  - } }}{x_{\scriptsize +} }-{k_{\scriptsize\overrightarrow { +  - } }}{\beta _{\scriptsize\overrightarrow { +  - } }}{x_{\scriptsize -} }}&0&{{k_{\scriptsize\overrightarrow { +  - } }}{\beta _{\scriptsize\overrightarrow { +  - } }}{x_{\scriptsize -} }}\\
				{{k_{\scriptsize\overrightarrow { -  + } }}{\beta _{\scriptsize\overrightarrow { -  + } }}{x_{\scriptsize +} }}&0&{1 - {k_{\scriptsize\overrightarrow { -  + } }}{\beta _{\scriptsize\overrightarrow { -  + } }}{x_{\scriptsize +} } - {k_{\scriptsize\overrightarrow { -  + } }}{\alpha _{\scriptsize\overrightarrow { -  + } }}{x_{\scriptsize -} } }&{{k_{\scriptsize\overrightarrow { -  + } }}{\alpha _{\scriptsize\overrightarrow { -  + } }}{x_{\scriptsize -} }}\\
				0&{{k_{\scriptsize\overrightarrow { -  - } }}{\beta _{\scriptsize\overrightarrow { -  - } }}{x_{\scriptsize +} }}&{{k_{\scriptsize\overrightarrow { -  - } }}{\alpha _{\scriptsize\overrightarrow { -  - } }}{x_{\scriptsize +} }}&{1 - {k_{\scriptsize\overrightarrow { -  - } }} + {k_{\scriptsize\overrightarrow { -  - } }}{x_{\scriptsize -} }}
				\end{array}} \right)}
		\end{array}}
	\end{array}\label{pingwenfenbu}
	\end{equation}
	%\end{small}
%\end{widetext}

The matrix $V$ is an approximation because it is possible that an individual reconnects its student set or teacher set of individuals. Since the population size is much larger than the average degree of the nodes, i.e., $N \gg L$, the approximation is completely acceptable.

The state space of the Markov chain is $S$. If ${k_{\scriptsize\overrightarrow { +  + } }}{k_{\scriptsize\overrightarrow { +  - } }}{k_{\scriptsize\overrightarrow { -  + } }}{k_{\scriptsize\overrightarrow { -  - } }}{x_ {\scriptsize +} }{x_ {\scriptsize -} } \ne 0$, there is a unique stationary distribution ${\pi_S} = \left( {{\pi _{\scriptsize\overrightarrow { +  + } }},{\pi _{\scriptsize\overrightarrow { +  - } }},{\pi _{\scriptsize\overrightarrow { -  + } }},{\pi _{\scriptsize\overrightarrow { -  - } }}} \right)$ determinded by equation ${\pi_S}Q = {\pi_S}$. We find that
%\begin{small}
\begin{equation}
\setcounter{equation}{4}
\renewcommand\theequation{A\arabic{equation}}
\label{eq.A.4}
\begin{array}{l}
{\pi_S} = {\cal N}\left( x_ {\scriptsize +} \right) * {\left[ {\begin{array}{*{20}{c}} {\displaystyle\frac{{x_ {\scriptsize + } ^2}}{{{k_{\scriptsize\overrightarrow { +  + } }}}}\left( {{x_ {\scriptsize +} }{\alpha _{\scriptsize\overrightarrow { +  - } }}{\beta _{\scriptsize\overrightarrow { -  + } }} + {{x_ {\scriptsize -} }\alpha _{\scriptsize\overrightarrow { -  - } }}{\beta _{\scriptsize\overrightarrow { +  - } }} + {{x_ {\scriptsize -} }\alpha _{\scriptsize\overrightarrow { -  + } }}\left( {{\alpha _{\scriptsize\overrightarrow { +  - } }} - {\alpha _{\scriptsize\overrightarrow { -  - } }}} \right)} \right)}\\
		{\displaystyle\frac{{{x_ {\scriptsize +} }{x_ {\scriptsize -} }}}{{{k_{\scriptsize\overrightarrow { +  - } }}}}\left( {{x_ {\scriptsize +} }{\alpha _{\scriptsize\overrightarrow { +  + } }}{\beta _{\scriptsize\overrightarrow { -  + } }} + {{x_ {\scriptsize -} }\alpha _{\scriptsize\overrightarrow { -  + } }}{\beta _{\scriptsize\overrightarrow { -  - } }}} \right)}\\
		{\displaystyle\frac{{{x_ {\scriptsize +} }{x_ {\scriptsize -} }}}{{{k_{\scriptsize\overrightarrow { -  + } }}}}\left( {{x_ {\scriptsize +} }{\alpha _{\scriptsize\overrightarrow { +  - } }}{\beta _{\scriptsize\overrightarrow { +  + } }} + {{x_ {\scriptsize -} }\alpha _{\scriptsize\overrightarrow { -  - } }}{\beta _{\scriptsize\overrightarrow { +  - } }}} \right)}\\
		\!\!{\displaystyle\frac{{x_ {\scriptsize - } ^2}}{{{k_{\scriptsize\overrightarrow { -  - } }}}}\left( {{{x_ {\scriptsize +} }\alpha _{\scriptsize\overrightarrow { +  + } }}{\beta _{\scriptsize\overrightarrow { -  + } }} + {{x_ {\scriptsize -} }\alpha _{\scriptsize\overrightarrow { -  + } }}{\beta _{\scriptsize\overrightarrow { +  - } }} + {{x_ {\scriptsize +} }\alpha _{\scriptsize\overrightarrow { +  - } }}\left( {{\alpha _{\scriptsize\overrightarrow { -  + } }} - {\alpha _{\scriptsize\overrightarrow { +  + } }}} \right)} \right)}
		\end{array}} \right]^\prime }
\end{array},
\end{equation}
%\end{small}
\noindent
where ${\cal N}\left( x \right) = \left[ {{{x_ {\scriptsize +} ^2\left( {{\alpha _{\scriptsize\overrightarrow { -  - } }}{\beta _{\scriptsize\overrightarrow { -  + } }}{x_ {\scriptsize -}} + {\alpha _{\scriptsize\overrightarrow { +  - } }}\left( {\left( {{\alpha _{\scriptsize\overrightarrow { -  + } }} - {\alpha _{\scriptsize\overrightarrow { -  - } }}} \right){x_ {\scriptsize -}} + {\beta _{\scriptsize\overrightarrow { -  + } }}{x_ {\scriptsize +}}} \right)} \right)} \mathord{\left/
			{\vphantom {{x_ + ^2\left( {{\alpha _{\scriptsize\overrightarrow { -  - } }}{\beta _{\scriptsize\overrightarrow { -  + } }}{x_ {\scriptsize -}} + {\alpha _{\scriptsize\overrightarrow { +  - } }}\left( {\left( {{\alpha _{\scriptsize\overrightarrow { -  + } }} - {\alpha _{\scriptsize\overrightarrow { -  - } }}} \right){x_ {\scriptsize -}} + {\beta _{\scriptsize\overrightarrow { -  + } }}{x_ {\scriptsize +}}} \right)} \right)} {{k_{\scriptsize\overrightarrow { +  + } }}}}} \right.
			\kern-\nulldelimiterspace} {{k_{\scriptsize\overrightarrow { +  + } }}}}} \right. + {x_ {\scriptsize +} }{x_ {\scriptsize -}}\left( {{\alpha _{\scriptsize\overrightarrow { +  + } }}{\beta _{\scriptsize\overrightarrow { -  + } }}{x_ {\scriptsize +}}} \right. + $
${{\left. {{\alpha _{\scriptsize\overrightarrow { -  + } }}{\beta _{\scriptsize\overrightarrow { -  - } }}{x_ {\scriptsize -}}} \right)} \mathord{\left/
		{\vphantom {{\left. {{\alpha _{\scriptsize\overrightarrow { -  + } }}{\beta _{\scriptsize\overrightarrow { -  - } }}{x_ {\scriptsize -}}} \right)} {{k_{\scriptsize\overrightarrow { +  - } }}}}} \right.
		\kern-\nulldelimiterspace} {{k_{\scriptsize\overrightarrow { +  - } }}}} + {{{x_ {\scriptsize +} }{x_ {\scriptsize -} }\left( {{\alpha _{\scriptsize\overrightarrow { +  - } }}{\beta _{\scriptsize\overrightarrow { +  + } }}{x_ {\scriptsize +}} + {\alpha _{\scriptsize\overrightarrow { -  - } }}{\beta _{\scriptsize\overrightarrow { +  - } }}{x_ {\scriptsize -}}} \right)} \mathord{\left/
		{\vphantom {{{x_ {\scriptsize +}}{x_ {\scriptsize -}}\left( {{\alpha _{\scriptsize\overrightarrow { +  - } }}{\beta _{\scriptsize\overrightarrow { +  + } }}{x_ {\scriptsize +}} + {\alpha _{\scriptsize\overrightarrow { -  - } }}{\beta _{\scriptsize\overrightarrow { +  - } }}{x_ {\scriptsize -} }} \right)} {{k_{\scriptsize\overrightarrow { -  + } }}}}} \right.
		\kern-\nulldelimiterspace} {{k_{\scriptsize\scriptsize\overrightarrow { -  + } }}}} + x_ {\scriptsize -} ^2\left( {{\alpha _{\scriptsize\scriptsize\overrightarrow { +  + } }}{\beta _{\scriptsize\scriptsize\overrightarrow { +  - } }}{x_ {\scriptsize +}} + } \right. {\alpha _{\scriptsize\overrightarrow { -  + } }}\left( {\left( {{\alpha _{\scriptsize\overrightarrow { +  - } }} - {\alpha _{\scriptsize\overrightarrow { +  + } }}} \right){x_ {\scriptsize +} }} \right.$
	${\left. {{{\left. { + {\beta _{\scriptsize\overrightarrow { +  - } }}{x_ {\scriptsize -} }} \right)} \mathord{\left/
				{\vphantom {{\left. { + {\beta _{\scriptsize\overrightarrow { +  - } }}{x_ {\scriptsize -} }} \right)} {{k_{\scriptsize\overrightarrow { -  - } }}}}} \right.
				\kern-\nulldelimiterspace} {{k_{\scriptsize\overrightarrow { -  - } }}}}} \right]^{ - 1}} > 0$ is a normalization factor. Here ${\pi _{\scriptsize\overrightarrow {XY} }}$ refers to the probability that a directed link $i$ is of type $\overrightarrow {XY} $ in the stationary regime.
\\

\section*{References}
%\bibliographystyle{1}
%\bibliography{IOPLaTeXGuidelines}
\bibliographystyle{unsrt}
\bibliography{IOPLaTeXGuidelines}

\end{document}